# Helices at Twisted Ferroelectric Interfaces

Kristina Holsgrove[1*], Ronan Lynch[1], Conor McCluskey[1], Eoin Moynihan[2], Andrew Rogers[1], Xinxin Hu[3], Boris Croes[4], Loan Hajdari-Manigault[4], Maria Lazarut[1], Eadaoin Anderson[1], Sam Seddon[5], Anna Razumnaya[6], Yurii Tikhonov[6], Ana M. Sanchez[2], Marios Hadjimichael[2], Amit Kumar[1], Raymond McQuaid[1], Jordi Arbiol[3,8], Gustau Catalan[3,8], Richard Beanland[2], Igor Luk'yanchuk[7], Lukas M. Eng[5,9], Salia Cherifi-Hertel[4], Tchavdar Todorov[1], J. Marty Gregg[1*]

1. Centre for Quantum Materials and Technologies, School of Mathematics and Physics, Queen's University Belfast, Belfast, BT71NN. U. K.
2. Department of Physics, University of Warwick, Coventry, U. K. CV4 7AL
3. Catalan Institute of Nanoscience and Nanotechnology - ICN2 (CSIC & BIST), Barcelona, Catalonia 08193, Spain
4. Université de Strasbourg, CNRS, IPCMS, UMR 7504, 67000, Strasbourg, France
5. TU Dresden, Institute of Applied Physics, Nöthnitzer Straße 61, 01187 Dresden, Germany
6. Jožef Stefan Institute, Jamova Cesta 39, Ljubljana 1000, Slovenia
7. University of Picardie, Laboratory of Condensed Matter Physics, Amiens 80039, France
8. Institució Catalana de Recerca i Estudis Avançats (ICREA), Barcelona 08010, Catalonia
9. Würzburg-Dresden Cluster of Excellence (EXC2147) ctd.qmat – Complexity, Topology, and Dynamics in Quantum Matter, Dresden, Germany

Communicating author emails: k.holsgrove@qub.ac.uk, m.gregg@qub.ac.uk

## Abstract

We report that twisted interfaces, formed between two ferroelectric lithium niobate single crystals, contain arrays of coherently bonded and elastically deformed nanoscale structural helices, separated from one another by screw dislocations. The helices generate remarkable contrast in four dimensional scanning transmission electron microscopy (4D STEM) and through related differential phase contrast (DPC). Such images have been successfully rationalised by modelling the trajectory of electrons that channel along helical pathways as they traverse the interface, but the structural helices implied have also been directly imaged using electron ptychography. Importantly, as lithium niobate is a uniaxial ferroelectric, the torsional strain in these structural helices necessarily also causes dipolar helices to form, inferred through consideration of pertinent piezoelectric tensors and flexoelectric coefficients. Helices are topologically distinct from the more commonly reported vortices and, importantly, they are chiral. Helical (and hence chiral) character was confirmed by observing that DPC images were identical before and after the sample was inverted. Chirality was also established through second harmonic optical circularly-polarised dichroism measurements, in which unusually large interfacial signals were seen. It must be stressed that the arrays of helices observed are distinct from moiré patterns; they facilitate lattice rotation combining plastic and elastic deformation and are not simply a superposition of lattices, twisted with respect to one another. Thus, intimately bonded twisted oxide interfaces will have distinctly different characteristics from those of twisted stacks of two-dimensional van der Waals materials, opening up a new and distinct kind of twistronics.

Ferroelectrics and ferromagnets are often seen as analogous, but they are clearly not the same. A major difference is in the way the dipoles interact and, specifically, in the anisotropy-to-exchange energy ratio, which is comparatively large for ferroelectrics. This results in dramatically thinner domain walls, for example, than typically seen in ferromagnets [1-4]. It has therefore been somewhat surprising to find that electrostatic analogues of complex magnetic topological patterns [5-7], involving extended regions of continuously changing dipole orientations, can not only be theoretically envisaged [8-12], but can also be widely experimentally realised in ferroelectrics [13-20]. In thin-film $PbTiO_3$-$SrTiO_3$ superlattices, for example, dipolar waves [13,14], vortices and vortex tubes [15-18], skyrmion bubbles [19] and merons [20] have all been reported.

Recently, widespread incorporation of water-soluble perovskite buffer layers into thin-film heterostructures has allowed free-standing membranes of functional oxides to be produced and investigated [21-23]. These have either been stacked on top of one another [24-28], as is commonly done with two-dimensional (2D) van-der-Waals (vdW) materials, or thermally bonded onto single crystal substrates [29,30]; notably, this has allowed the effects and impact of relative twists between stacked layers to be probed. Twist-induced strain gradients have been seen to generate topological complexity, with flexoelectrically-induced vortex arrays found in a variety of twisted perovskite bilayer systems: ferroelectric $BaTiO_3$ [25]; incipient ferroelectric $SrTiO_3$ [26, 27]; and even non-ferroelectric, conducting $SrRuO_3$ [28]. Thus, both epitaxial superlattices and twisted oxide membranes host novel and complex electrical dipolar patterns.

However, thin film fabrication techniques are not a prerequisite for this kind of research. Twisted interfaces can also meaningfully be produced through straightforward bonding of single crystals. Indeed, the robust nature of bulk material makes high temperature thermo-compressive processing straightforward and can ensure that the resulting intimately bonded interfaces go well beyond vdW interactions. The approach has recently been used to reveal surprising effects of twist on the emergence and suppression of interfacial conductivity in oxide ferroelectrics [31]. The almost ideal nature of the single crystals and their bonded interfaces enables very precise studies of the lattice distortions induced by twisting. This is important because the change from the localised strain fields associated with dislocations to the continuous strain gradients reported in twisted bilayers remains obscure, as is the impact of that transition on polar topology.

In this article, we report that thermo-compressively bonded twisted lithium niobate (LNO) single crystals generate topologically non-trivial structures at, and close to, their shared interface. Relative lattice twist is predominantly facilitated through the formation of a network of screw dislocations, as expected classically [29,30,32-34]. However, elastic torsional strain is also evident, manifest most prominently in the formation of arrays of nanoscale structural helices, with periodicities in the ~10nm range. Piezoelectricity and flexoelectricity demand that helical distortions in the lattice are accompanied by dipolar patterns with the same chirality. Thus, both structural and dipolar helices coexist. Direct evidence of chirality, found by imaging before and after 180° sample rotation and through the existence of strong second harmonic optical circularly-polarised dichroism, helps confirm the helical topology.

**Bonding the Crystals and Imaging the Interface**

Bonded interfaces were made by placing two pieces of clean z-cut LNO single crystal on top of one another, with a relative twist in the orientations of their primary *a*-*b* crystallographic axes. Polarisation axes (*c*-axes) were oriented either in the same sense (creating “neutral” interfaces), or with opposing senses (creating “head-to-head” or “tail-to-tail” interfaces with significant bound charge densities). Pressure was applied and bonding was achieved by sintering in a conventional chamber furnace. Details can be found in reference [31].

Once bonded, cross-sections were cut and polished to allow focused ion beam milling of thin electron-transparent lamellae, incorporating the interface, with lamellar planar normals at a slight angle to the crystallographic *c*-axes of the LNO (fig. 1a). Interfaces could then be imaged in plan-view, using Transmission Electron Microscopy (TEM), with the electron beam aligned along the $[0001]/[000\bar{1}]$ trigonal zone axes of the bonded single crystals. Electrical dipolar components, perpendicular to the electron beam, are known to generate signal in Differential Phase Contrast (DPC) and four-Dimensional Scanning Transmission Electron Microscopy (4D STEM) [33]. In this context, the *c*-axis beam propagation direction was crucial, insofar as it is oriented parallel to the spontaneous polarisation in the uniaxial bulk crystals. Conventionally oriented ferroelectric polarisation (aligned || to the z-axes and hence || to the incident electron beam) therefore had no influence on the observed DPC or 4D STEM contrast (no bulk polar components lay perpendicular to the zone axis) and so information gained could confidently be attributed to non-axial features associated with the interface itself.

As can be seen from conventional high-resolution TEM (fig. 1b), the microstructure has periodically distributed bright regions, in which lattice fringe and atomic column contrasts are clear. Such regions are separated from one another by darker contrast boundaries in which lattice fringes are less strongly evident. Weak beam dark-field imaging (fig. 1c and supplementary information S1) reveals the darker regions to be associated with a network of screw dislocations, which facilitate relative interfacial lattice twist. The geometrical topology of these dislocation networks and estimates of their characteristic periodicities are described in more detail in section S2 of the supplementary information.

Unlike results from recent research on twisted $SrTiO_3$ membranes [26], the twisted LNO bi-crystal interfaces show striking STEM DPC contrast within the coherently bonded regions (hexagonal regions bound by the screw dislocations) in which the motion of the centre of mass of the electron diffraction intensity in the back focal plane, recorded on an annular quadrant detector, clearly suggests vorticity (figs. 1d & 1e). Reconstructing the in-plane DPC vector field, by interpolating the measured data and tracing their integral curves, provides interesting and additional detail regarding topology (supplementary information S3).

A similar contrast is observed using 4D STEM imaging (supplementary information fig. S4.1). Here, the migration of the centre of the observed intensity in the straight-through beam and across all the first-order diffraction spots in the diffraction pattern, can be separately analysed using diffraction filters, as a function of position on the specimen (fig. S4.1a & c). The senses of rotation within the vortices, implied using the bright-field and dark-field filters, are opposite in nature. This might relate to electric field versus polarisation information [35], but we are conscious that the interpretations used in previous publications may not be universally true.

Data presented in fig. 2a (and in supplementary fig. S5.1) and comparisons with calculated elastic strain maps (supplementary information S6) help progress our understanding, insofar as they make it clear that the vorticity contrast seen in both DPC and 4D STEM superposes onto localised rotational strain. Vorticity and elastic torsional strain hence spatially coexist. Elastic accommodation of twist over a finite length-scale implies the existence of structural helices which were directly confirmed, using TEM ptychography, by tracking the changes in niobium atomic positions in sequential ptychographic planes (in 4Å steps) on approaching and passing through the bonded interface (supplementary figures S7.1-S7.3 and the supplementary video SV.1). It is interesting that the helical pitch reaches a distinct minimum (torsional distortion is maximum) at the interface itself (figure S7.4).

Torsional elastic strain necessarily generates unit cell distortion and tilting of the bonds that would have been aligned parallel to the *c*-axis in the undistorted LNO crystal (presented schematically in fig. 2b).

We have modelled electron propagation tracks (current pathways) for a beam of electrons, initially incident perpendicular to the interface, as would be the case for STEM imaging (see supplementary information S8). Such modelling reveals that electrons channel along pathways in the material with high electron density such that, within the interfacial region, currents track the helical locus of the tilted bonds referred to above and thereby enter and exit the lamellar sample with a circumferential displacement (fig. 2c). Thus, elastic torsional strain (seen through modelling and STEM analysis) and structural helices (confirmed by TEM ptychography) are entirely consistent with the vortical electron propagation contrast generated in DPC and 4D STEM imaging.

Since all ferroelectrics are piezoelectric and all materials are flexoelectric, helical strain implies that polarisation components beyond the conventional *c*-axis spontaneous polarisation must be present. We have used the piezoelectric tensor and an order of magnitude effective flexoelectric coefficient for LNO to determine them (see supplementary information S9). Piezoelectricity generates a largely toroidal polarisation component in the plane perpendicular to the torsion axis (fig. 2d), which rotates on moving away from the interface in concert with the structural helical rotation. Flexoelectricity generates a somewhat similar polarisation contribution, but in an opposing sense and smaller by an order of magnitude (fig. 2d). The combined result is that, in the elastically twisted regions, additional dipolar patterns emerge that are dominated by piezoelectricity. Superposition of the large spontaneous polarisation aligned along the crystallographic *c*-axes leads to the complex largely helical pattern in the dipoles. It is noteworthy that, even in tail-to-tail or head-to-head interfaces, the handedness of structural distortions and dipolar patterns is the same.

**Confirmation of Interfacial Chirality**

Defining vortices as either clockwise or anticlockwise has little meaning, as this classification is subject to change, if the vortex orientation alters with respect to an observer. Unlike vortices, helices are genuinely chiral and maintain a distinct handedness that is independent of viewing direction. To distinguish between vorticity and helicity at the twisted ferroelectric interfaces, we therefore examined and compared plan-view DPC images with incident electron beams parallel to the $[0001]$ and the $[000\bar{1}]$ trigonal zone axes of one of the bonded LNO crystals. Put plainly, we took plan-view DPC images of the interface, turned the sample upside-down, imaged once again, and then compared the two results. Panels a to f in fig. 3 summarise observations made. Most notably, the DPC contrast (panels c and f) is found to be insensitive to the viewing direction. This would be impossible if the contrast origin were related to vortices, but perfectly consistent with helices, which in this case would have right-handed chirality.

Importantly, collections of individual helices across extended regions were found to have the same handedness (see fig. 1d for example). Mean-field properties associated with the interface as a whole might therefore also be expected to be chiral. This was largely confirmed through optical investigations (see supplementary information S10): left or right-handed circularly-polarised light was used in an incident confocal beam to excite second harmonic signals from different voxel locations in both the bulk LNO and at the twisted interface. As expected, second harmonic generation (SHG) was primarily associated with the interface. Moreover, there was an obvious difference in the intensity of this interfacial SHG, depending on whether right or left-handed circularly polarised incident light was used (fig. 3g). Subtracting the two intensities measured at each voxel allowed spatial mapping of the optical dichroism and hence any variations in the chiral nature of the interface. In fig. 3h, one can see a plan-view representation of chirality across a region of the interface in which most of it is right-handed. However, small patches in which the chirality reverses can also be seen.

**Helical Arrays and Moiré Patterns**

Our observations demonstrate that the interfacial microstructure in the twisted bonded LNO crystals is generally characterised by a hexagonal network of screw dislocations separating helical elastically deformed regions which contain piezoelectrically and flexoelectrically induced dipolar helices. The combination of plastic and elastic deformation is qualitatively different from the direct superposition of two undistorted lattices and resulting moiré patterns that are characteristic of twisted interfaces in stacked vdW materials. However, the initial state prior to the microstructural evolution that occurs during thermo-compressive synthesis, is indeed moiré (the two crystals were initially simply pressed on top of each other at room temperature). A link between the initial moiré period and the evolved helical arrays should therefore exist, as symmetry-related points in the moiré pattern might be expected to change identically at elevated temperature.

For twist angles close to 0° and 60° (where a perfect coincident lattice mesh exists for shared (0001) interfacial planes in the trigonal system), this seems to be true. In fig. 4a and 4b, for example, it is clear that the superposition of two LNO structures with a relative twist of 56° generates effective moiré periodicities that are comparable to the periodicities of the helices (~5nm). At 21°, the helical arrays revealed by DPC have a period of approximately 15 nm (fig. 4e), much larger than the conventional atomic moiré period of approximately 0.8 nm observed by HRTEM (figs. 4c,d). This difference is consistent with the interface lying close to the $\Sigma 7$ coincidence-site-lattice (CSL) rotation angle of 21.7868° (see supplementary information S2 for more detail). The deviation from this angle, $\Delta\theta = |21° - \theta\Sigma 7| \approx 0.79°$, gives rise to a secondary dislocation network. Using the shortest Displacement Shift Complete (DSC) Burgers vector, $\boldsymbol{b}_{\mathrm{DSC}} = a/\sqrt{7} \approx 0.195$ nm, the estimated spacing of this network is $L \approx b_DSC/\Delta\theta \approx 14$ nm, close to the observed helix array period.

## Outlook

It is clear that bonded functional oxides adapt to imposed twist in ways that would be unexpected for stacked and twisted vdW systems. There is hence a great opportunity for new discovery and this is accentuated by the rather diverse observations made so far: several studies, in which individual functional oxide membranes are twisted and stacked, report continuous modulations in elastic strain and dipolar patterns dominated by flexoelectricity [25,27,28]; others find the formation of dislocation networks which accommodate all the relative twist, leaving significant regions between dislocations largely free from elastic deformation and induced polarisation [26,29,30]. In our study, we find a combination of these, with dislocation networks and localised helical elastic deformation leading to strong chirality and the inferred existence of dipolar helices. A rich behavioural landscape therefore clearly exists.

Creating and manipulating dipolar topological objects is not just of academic interest. An exciting strand in data-related research has been in the use of ferromagnetic skyrmions as information carriers [36-39], made to travel along race-tracks. Demonstrator designs have already been shown capable of reservoir computation [39]. Importantly, magnetic skyrmions are driven by currents. An equivalent technology, using current-free voltage control to manipulate electrical dipolar topological objects, might be attractive and has indeed recently been theoretically considered [40]. In addition, the use of designed twist-defined interfaces could be useful for specific chiral optical or phononic applications.

**References:**

[1] V. A. Zhirnov, Sov. Phys. JETP **35**, 822 (1959)
[2] S. P. Ray and D. I. Paul, J. Appl. Phys. **41**, 5238 (1970)
[3] C. Kittel, Solid State Communications **10**, 119 (1972)
[4] G. Catalan *et al.*, J. Phys: Cond. Matter, **19**, 022201 (2007)
[5] T. Shinjo *et al.*, Science **289**, 930 (2000)
[6] S. Muhlbauer *et al.*, Science **323**, 915 (2009)
[7] N. Nagaosa and Y. Tokura, Nat. Nanotech. **8**, 899 (2013)
[8] I. I. Naumov, L. Bellaiche and H. Fu, Nature **432**, 737 (2004)
[9] Y. Nahas *et al.*, Nat. Commun. **6**, 8542 (2015)
[10] I. Luk'yanchuk *et al.*, Nat. Commun. **11**, 2433 (2020)
[11] S. Yuan *et al.*, Phys. Rev. Lett. **130**, 226801 (2023)
[12] M. A. P. Goncalves *et al.*, Sci. Adv. **5**, eaau7023 (2019)
[13] P. Tong *et al.*, Nat. Phys. **21**, 464 (2025)
[14] J. M. Gregg, Nat. Phys. **21**, 344 (2025)
[15] A. K. Yadav *et al.*, Nature **530**, 198 (2016)
[16] V. A. Stoica *et al.*, Nat. Mater. **18**, 377 (2019)
[17] E. Zatterin *et al.*, Phys. Rev. X **14**, 041052 (2024)
[18] D. Rusu *et al.*, Nature **602**, 240 (2022)
[19] S. Das *et al.*, Nature **568**, 368 (2019)
[20] Y. J. Wang *et al.*, Nat. Mater. **19**, 881 (2020)
[21] D. Lu *et al.*, Nat. Mater. **15**, 1255 (2016)
[22] S. S. Hong *et al.*, Science **368**, 71 (2020)
[23] G. Dong *et al.*, Science **366**, 475 (2019)
[24] M. G. Hernandez *et al.*, Adv. Phys. Res. **5**, e00219 (2026)
[25] G. Sanchez-Santolino *et al.*, Nature **626**, 529 (2024)
[26] W. Sandholt *et al.*, Designing dislocation-driven polar vortex networks in twisted perovskites arXiv:2603.27272
[27] I. Tenreiro *et al.*, Adv. Mater. e12969 (2026)
[28] Y. Lun *et al.*, Polarization vortices in a ferromagnetic metal via twistronics. Nat. Mater. (2026). https://doi.org/10.1038/s41563-026-02755-8
[29] M. Schmidbauer *et al.*, Appl. Phys. Lett. **126**, 101902 (2025)
[30] J. Maltitz *et al.*, Phys. Rev. B **113**, 214110 (2026)
[31] A. Rogers *et al.*, Nat. Commun. **17**, 1842 (2026)
[32] F. C. Frank, Conf. Plastic Deformation of Crystalline Solids, Melon Institute, Pittsburgh (1950) p.150
[33] W. T. Read, *Dislocations in Crystals.* McGraw-Hill, New York (1953)
[34] R. C. Pond and D. S. Vlachavas, Proc. Roy. Soc. A **386** 95-143 (1983)
[35] C. Addiego *et al.*, APL Mater. **11**, 051106 (2023)
[36] S. Woo *et al.*, Nat. Mater. **15**, 501 (2016)
[37] B. He *et al.*, Nano Lett. **23**, 9492 (2023)
[38] A. Fert *et al.*, Nat. Nanotech. **8**, 152 (2013)
[39] K. Raab *et al.*, Nat. Commun. **13**, 6982 (2022)
[40] J. Iniguez-Gonzalez and H. Aramberri, Creating currents of electric bubbles. https://doi.org/10.48550/arXiv.2412.15074
[41] Z. Chen *et al.* Science **372**, 826–831 (2021)
[42] C. H. Lee, Microsc. Microanal. **31**, ozaf070 (2025)
[43] McEniry *et al.* J. Phys.: Condens. Matter **19**, 196201 (2007)

[44] Dataset for “Periodic Arrays of Nanoscale Helices at Twisted Ferroelectric Interfaces”, Queen’s Research Repository (2026)

## Methods

Sample Fabrication

A FIB microscope was used to prepare cross-sectional lamellae for transmission electron microscopy (Tescan-focused ion beam secondary electron microscope Lyra 3). FIB milling was performed with gallium ions at accelerating voltages of 30 kV (roughmilling), 5, 2, and 1 kV (fine polishing).

Transmission Electron Microscopy

STEM measurements were acquired on a ThermoFisher Scientific Talos F200X instrument operated at a 200kV accelerating voltage. For DPC imaging, a segmented detector (ThermoFisher) was used to measure deflection of the transmitted electron beam. 4DSTEM was acquired on a double-corrected JEOL ARM200F operating at 200 kV, in Nanobeam STEM mode, with a 3 mrad semi-convergence angle and equipped with a Quantum Detectors Merlin pixelated detector. Conventional TEM imaging and selected area diffraction was performed using a JEOL 2100plus HC $LaB_6$ microscope at 200 kV.

Multislice Electron Ptychography

For ptychography, the 4DSTEM experiments were performed using a double aberration-corrected Spectra 300 microscope operated at an accelerating voltage of 300 kV. The 4DSTEM datasets were acquired using an electron microscope pixel array detector (EMPAD) with a probe convergence semi-angle of 20.2 mrad. The electron probe was overfocused by approximately 15 nm relative to the specimen entrance surface and raster-scanned with a step size of 0.37 Å [41]. At each scan position, a 128 × 128 pixel diffraction pattern was recorded with a dwell time of 1 ms and a beam current of approximately 35 pA, corresponding to an electron dose of approximately $1.6 \times 10^6$ $e^- Å^{-2}$. Multislice ptychographic reconstructions were performed using PtyRAD [42], employing five probe modes, a slice thickness of 0.4 nm, and more than 1,000 iterations. Atomic column positions were extracted from the reconstructed phase images using Atomap. Nb columns were matched between consecutive slices, with the slice sequence traversed in opposite directions for the upper and lower layers. For each matched column, the in-plane displacement was defined as its position in the target slice minus its position in the reference slice. The mean displacement vector of all matched columns in each slice pair was then subtracted to obtain the relative displacement field.

Modelling of Current Propagation Pathways

The transport simulations down twisted multilayers employ the Hairy Probes method for electronic open boundaries [43]. The electronic structure is described through a single-orbital tight-binding model with hopping amplitudes that are decaying functions of interatomic separation (for details, see Supplementary Material S4). This is the minimal model required to capture the connection between the real-space current-flow geometry and the underlying atomistic structure. The description of transport it provides is generic but has been tested in numerous past studies and has always been found to give a satisfactory qualitative picture of the physics involved.

Optical Circular Dichroic Second Harmonic Generation

Second-harmonic generation (SHG) experiments are conducted using a mode-locked Ti:sapphire oscillator delivering 100 fs laser pulses at a central wavelength of 800 nm and a repetition rate of 80 MHz in an inverted optical microscope, with detection in a back-reflection geometry as described in Ref. [https://doi.org/10.1063/5.0037286]. The optical axis is aligned normal to the sample surface and parallel to the polar axis. The laser beam is focused into the volume of the twisted crystals through a 60x objective with a numerical aperture of 0.85. Three-dimensional SHG imaging is performed by translating the sample relative to the fixed laser focus along the x, y, and z directions, thereby scanning the excitation voxel laterally across and axially through the twisted-crystal interface. The polarisation state of the incident light is controlled using a combination of achromatic half-wave and quarter-wave plates, allowing switching between left-circularly polarised (LCP) and right-circularly polarised (RCP) excitation. The wave plates are adjusted to compensate for polarisation distortions introduced by the optical system and to maximise the degree of circular polarisation at the sample. For all circular dichroic (CD)-SHG measurements, an ellipticity ratio above 0.95 is maintained for both LCP and RCP excitation. CD-SHG images are acquired sequentially under LCP and RCP excitation while the sample is raster scanned with a spatial step of 50 nm and a pixel dwell time of 20 ms. The CD-SHG contrast is quantified from the normalised difference between the SHG intensities measured under LCP and RCP excitation.

**Data Availability**
The data generated in this study have been deposited on the Queen's Research Repository database [44].

**Author Contributions**
Conceptualization: JMG, LME, GC, RB, SC-H, IL, TT
Methodology: KH, RL, CMcC, EM, AR, XH, BC, LH-M, AK, RMcQ, JA, RB, IL, SC-H, TT, JMG
Investigation: KH, RL, CMcC, EM, AR, XH, BC, LH-M, ML, EA, SS, AR, YT, AMS, MH, JA, GC, RB, IL, SC-H, TT, JMG
Visualization: KH, RL, CMcC, EM, XH, BC, LH-M, ML, AR, YT, IL, SC-H
Funding acquisition: JMG, LME, AK, GC, JA
Project administration: JMG, LME
Supervision: JMG, AS, IL, LME, SC-H, GC, JA
Writing – original draft: KH, RL, CMcC, RB, IL, SC-H, TT, JMG
Writing – review & editing: KH, RL, CMcC, SS, MH, IL, GC, LME, SC-H, TT, JMG

**Competing interests:** All authors declare that they have no competing interests.

**Funding and Acknowledgements**
JMG acknowledges financial support from the UK Engineering and Physical Sciences Research Council (EP/X027074/; JMG) for the CAMIE Programme Grant. KH acknowledges support from the Department of Education and Learning NI through grant USI-205; AK and AR acknowledge financial support from the UK Department for the Economy (DfE). RMcQ gratefully acknowledges support by a UKRI Future Leaders Fellowship (Grant Number MR/T043172/1; R.G.P.M.). SC-H and LH-M acknowledge funding from the Interdisciplinary Thematic Institute QMat (ITI 2021–2028) and the EUR QMat (ANR-17-EURE-0024), supported by the IdEx Unistra (ANR-10-IDEX-0002) and SFRI STRAT'US (ANR-20-SFRI-0012) through the French Programme d'Investissements d'Avenir. LME and SS gratefully acknowledge financial support by the Deutsche Forschungsgemeinschaft (DFG) through the DFG research group FOR5044 (grant no. 426703838; https://www.for5044.de). Furthermore, LME also gratefully acknowledges the financial support by the Deutsche Forschungsgemeinschaft (DFG) through the Würzburg-Dresden Clusters of Excellence ct.qmat – Complexity and Topology in Quantum Matter, and ctd.qmat – Complexity, Topology, and Dynamics in Quantum Matter (EXC 2147, project-id 390858490). GC acknowledges the EU's HORIZON--MSCA Staff Exchange Action 3D-TOPO and Spain's AEI project PID2023-148673NB-I00 (Project GRIPHO2). IL, AR and GC acknowledge the HORIZON-MSCA-SE-3D-TOPO action (GA-101236483). AR and YT acknowledge the support from the Slovenian Research and Innovation Agency (projects J1-70018, 5110-18/2025-5) and the HORIZON-MSCA-COFUND-SQUASH project (GA-101177446). ML and TT thank Professor Hugo van der Hart for helpful discussions. We also acknowledge ICN2 and the Joint Electron Microscopy Center at ALBA (JEMCA) for providing key facilities. ICN2 is supported by the Severo Ochoa program from Spanish MCIN / AEI (Grant No. CEX2021-001214-S), the European Regional Development Fund from European Union (Grant No. IU16-014206 (METCAM-FIB)), and the CERCA Programme, Generalitat de Catalunya. ICN2 is a founding member of e-DREAM. JA and XH acknowledge funding from Generalitat de Catalunya (Grant No. 2021SGR00457). This study is part of the Advanced Materials programme and was supported by MCIN with funding from European Union

NextGenerationEU (PRTR-C17.I1) and by Generalitat de Catalunya (In-CAEM Project). ICN2 acknowledges funding from Grant IU16-014206 (METCAM-FIB) funded by the European Union through the European Regional Development Fund (ERDF), with the support of the Ministry of Research and Universities, Generalitat de Catalunya. X.H. acknowledges PhD scholarship support from the China Scholarship Council (CSC) (Grant No. 202304910019), and support from the Severo Ochoa Centre of Excellence programme (Grant No. CEX2021-001214-S/MICIU/AEI/10.13039/501100011033). The authors acknowledge the use of the Spectra 300 microscope, provided under ALBA Synchrotron proposal (Grant No. 20260390006). MH acknowledges financial support from the Engineering and Physical Sciences Research Council (EPSRC) under Award no. UKRI2034.

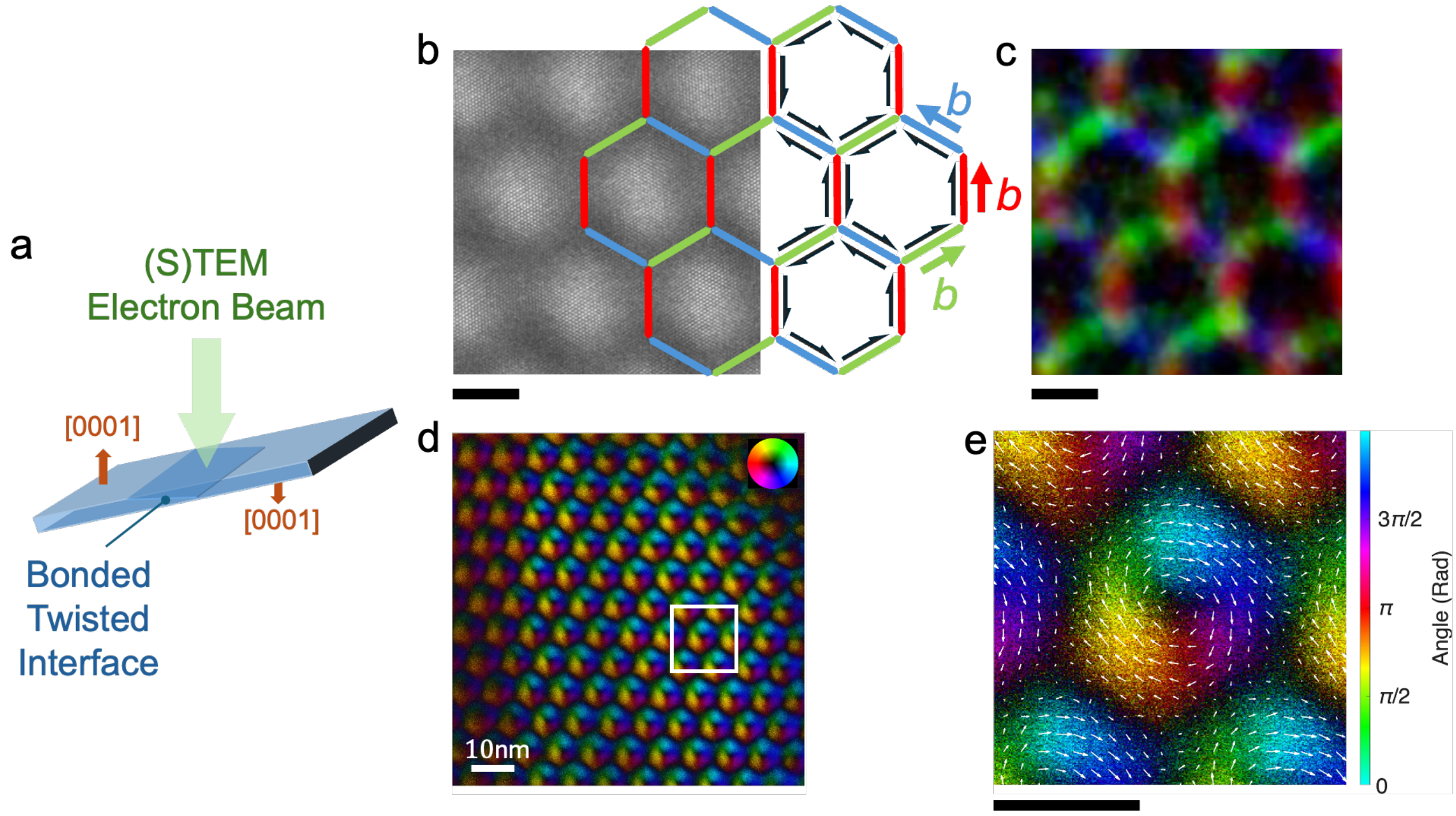

**Figure 1: Plan-view imaging of twisted lithium niobate interfaces using transmission electron microscopy (TEM) |** Focused ion beam milling was used to cut electron transparent lamellae at modest inclination angles with respect to the twisted thermo-compressively bonded interfaces. Individual lamellae were subsequently oriented for TEM investigation with the incident electron beam aligned parallel to the trigonal c-axes of the bonded single crystals and hence perpendicular to the interface itself (a). In this schematic a charged head-to-head interface is shown, as is pertinent for the rest of the panels. However, the same kind of TEM images were also observed in tail-to-tail and head-to-tail polar interfaces. High-resolution parallel beam imaging (b) showed a periodic array of regions of bright contrast in which lattice fringes were clear, surrounded by a hexagonal network of darker regions without obvious lattice fringe contrast. Weak-beam dark field imaging (c) showed this network to be associated with dislocations in which lines and Burgers vectors were evidently parallel (screw dislocations), as indicated by the superposed schematic in (b). Dark grey half-arrows in this schematic indicate the relative shear associated with the screw dislocations, which enable twist across the interface. As shown in (d) and (e), differential phase contrast scanning transmission electron microscopy (DPC STEM), using a quadrant annular detector, produced striking images showing apparent vorticity within the coherently bonded, dislocation-free, regions. Scale bars in panels b, c and e are 5 nm in length.

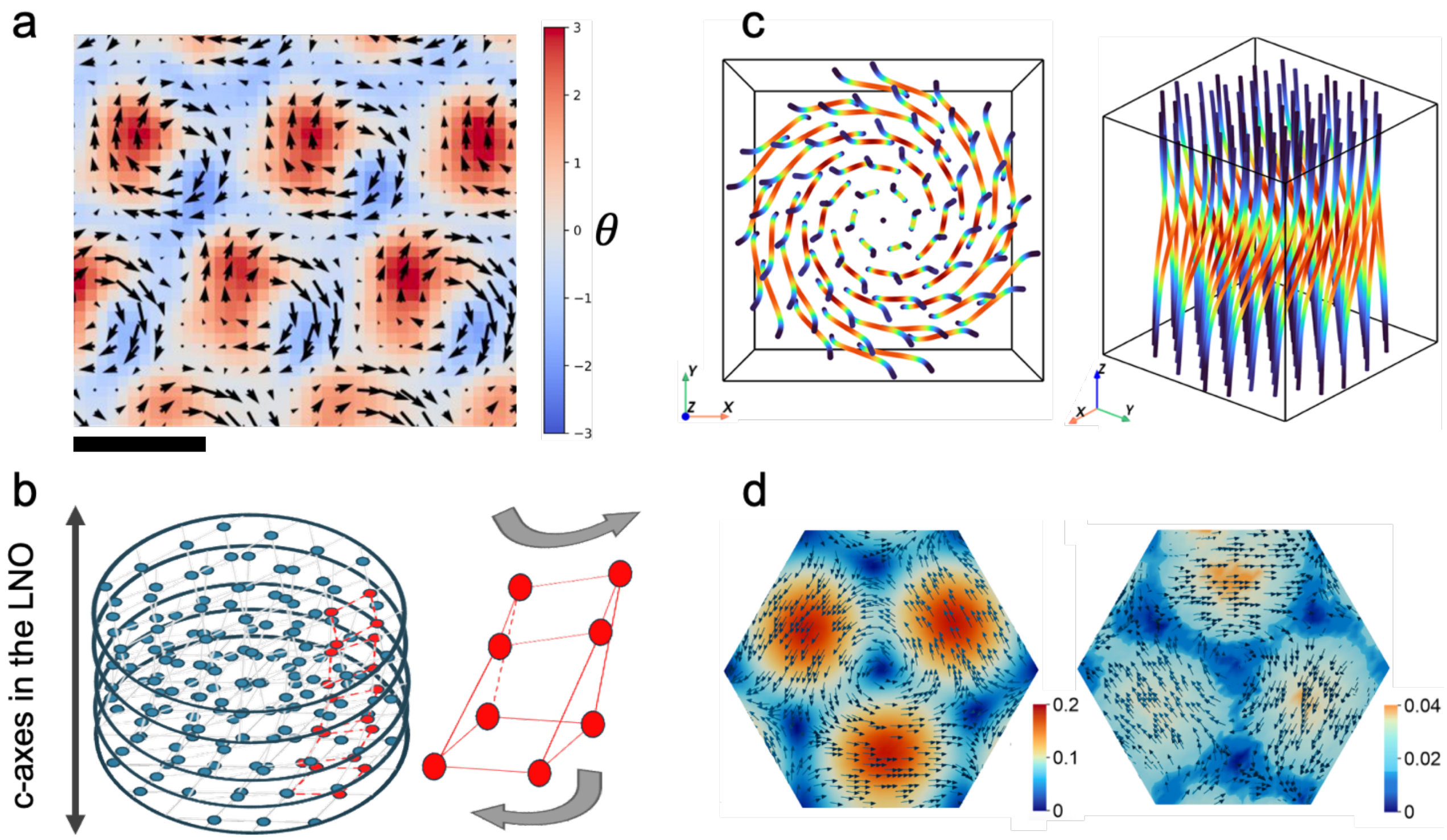

**Figure 2: Helices in both structure and dipolar configurations |** Superposing vorticity (from four-dimensional scanning transmission electron microscopy (4D STEM)) and lattice distortion information (from STEM strain analysis) shows obvious spatial coincidence (a). In this case the rigid lattice rotation representation is given, as this best reveals the link between 4D STEM vortical contrast and torsional elastic lattice deformation. Elastic twisting generates structural helices and associated torsional distortion in the LNO unit cells at and close to the bonded interface (schematically shown by the twisted lattice in (b)). Bonds that were vertical in the undistorted lattice, now locally tilt. Electron propagation pathways for electrons incident in STEM imaging have been modelled. They tend to follow the bonds that are most closely aligned with their incident direction and so channel along the tilted bonds in the interface region. Pathways hence also show a helical form. The left-hand panel in (c) projects streamlines associated with modelled electron propagation pathways through the interface, viewed along the incident beam direction. Colours indicate local streamline orientations relative to “x” (orange), “y” (green) and “z” (blue). Circumferential shifts between interface entry and exit points can be interpreted for each streamline from the separation between the blue sections on each streamline. The right-hand panel shows the same information, using a different projection. Modelling thus shows that imaged “vorticity” in STEM is entirely commensurate with the existence of structural helices. It should be noted that panels in (c) are associated with an exaggerated torsional strain and should be taken as indicative only. Helical elastic strain induces piezoelectric and flexoelectric dipolar contributions, which augment the spontaneous ferroelectric polarisation in the LNO single crystals. These have been calculated in the unit cell adjacent to and above a head-to-head charged interface (see supplementary information S9). Both piezoelectrically induced polarisation (d - left panel) and flexoelectrically induced polarisation (d - right panel), within the hexagonal plate regions bounded by screw dislocations, are broadly vortical (helical in three dimensions). Black arrows indicate local in-plane dipolar directions, while the colours indicate polar magnitudes. It can be seen that the sense of the piezoelectrically and flexoelectrically induced polarisations broadly oppose but that piezoelectricity dominates. Scale bar in a is 5nm.

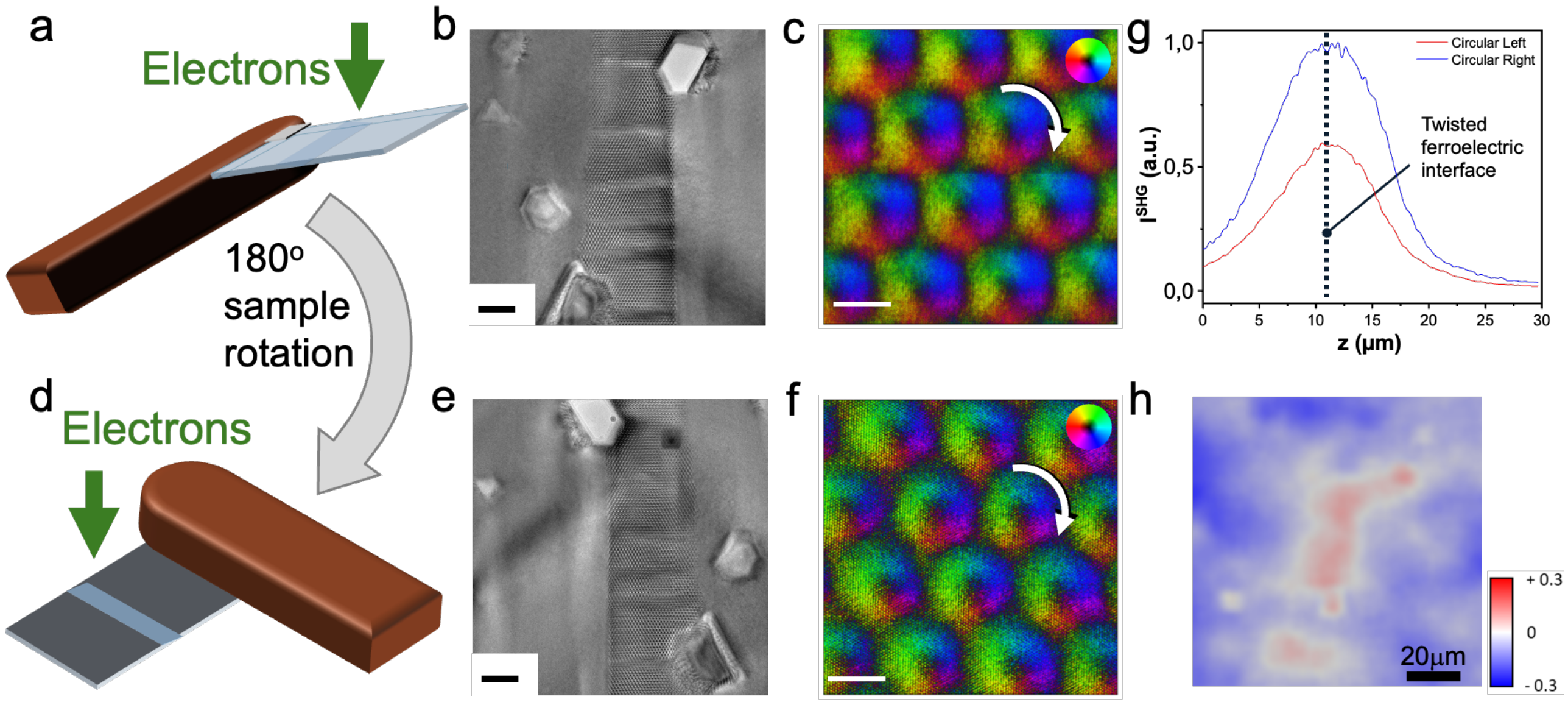

**Figure 3: Confirming helicity in the elastically bonded regions of the twisted interface |** Investigation of a twisted bonded interface prior to physical inversion of the entire sample (schematically represented in panel (a)) shows the stripe region in which the interface has been captured within a focused ion beam cut lamella (b); this stripe shows obvious periodic mottled contrast. Note the holes and damage distributed across the image. DPC STEM imaging (c) reveals that this mottled interfacial contrast is associated with periodic vorticity, as seen before. In this instance it has a clockwise sense. When the sample is inverted (d), the physical reorientation becomes apparent in the altered relative positions of holes and damage in transmission imaging (e). DPC STEM again shows that the interface has periodic vorticity (f). Crucially, despite the sample inversion, the sense of the vorticity is still clockwise. This can only be possible if the vortices are actually helices, with a handedness that is insensitive to orientation. The handedness of individual nanoscale helices, within the DPC imaging areas, all appear to be the same and hence mean-field chiral behaviours might be expected. Circularly polarised optical second harmonic imaging shows this to be true, insofar as the second harmonic intensity generated by right-handed circularly polarised incident light at the twisted interface (g) is significantly more intense than that generated from left-handed circularly polarised light. Subtracting the two signals allows the dichroism to be determined and its variation mapped across large regions of the twisted interface (h). This plan-view image shows that, while most of the interface has right-handed chirality (given by blue contrast), there are some patches in which the handedness changes (in red). Given a well-defined relative orientational twist between the bonded crystals, this lack of uniformity in the local chirality is both surprising and interesting. Scale bars in panels b and e are 100nm; in c and f they are 5 nm.

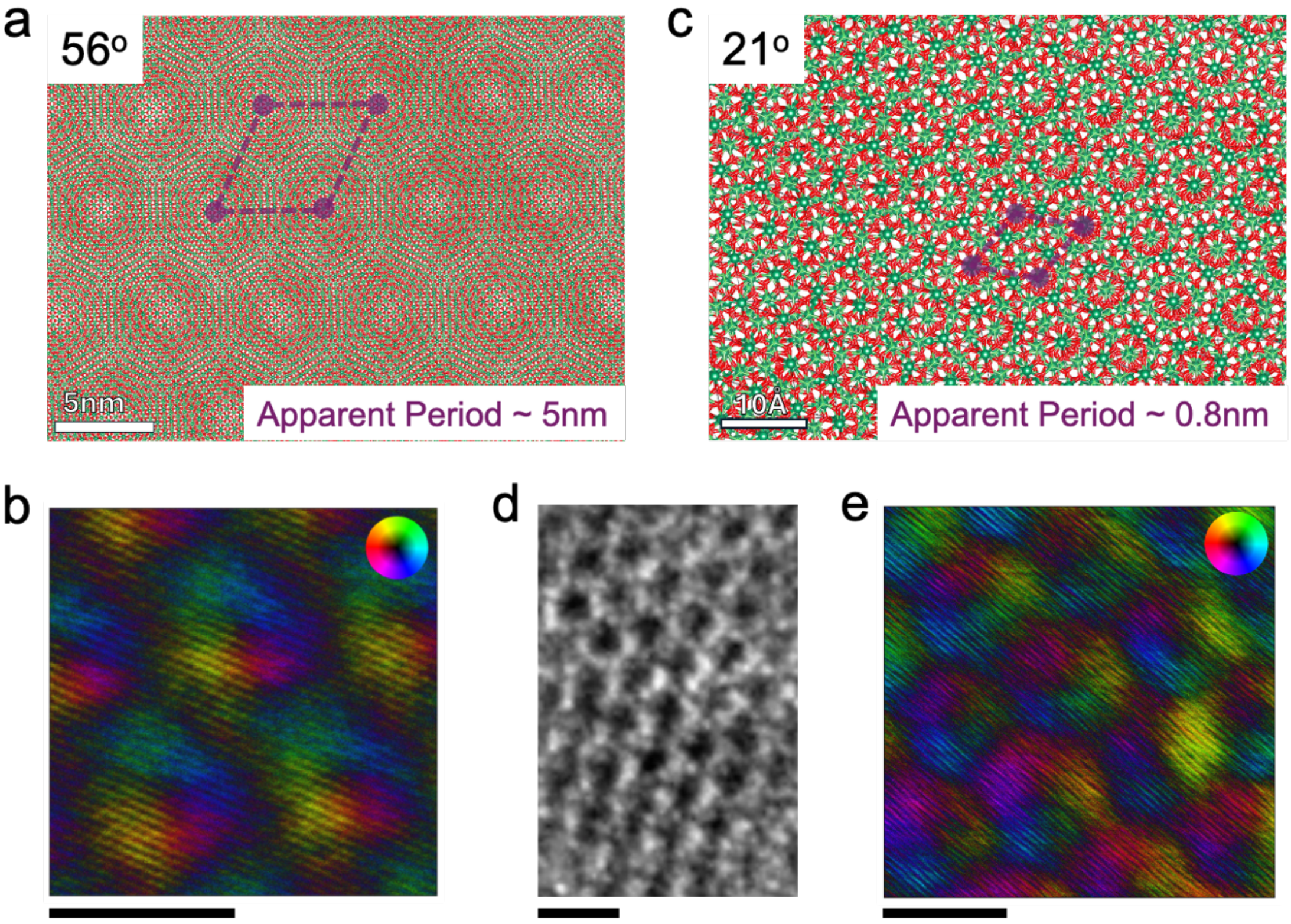

**Figure 4: Comparison of moiré and helix periods |** At an angular twist of 56° (4° away from that of a perfect interfacial coincident site lattice that exists at 60°), direct superposition of two LNO layers yields a moiré pattern with an apparent periodicity of ~5 nm, highlighted by the purple unit cell overlay (a). This correlates well with the periodicity measured from DPC STEM (b). For a 21° relative twist, on the other hand, the expected moiré period of ~0.8 nm (c), while seen in high resolution TEM (d), is over an order of magnitude smaller than the structural and dipolar helices revealed by DPC STEM (e). Panels a and c were rendered using Vesta with Nb atoms shown in green. Scale bars on panels b, d and e are 5 nm, 1 nm and 15 nm, respectively.

# Supplementary Information

## S1: Weak Beam Dark Field (WBDF) Imaging

To be unaffected by the strain field due to a dislocation, of Burgers vector $\boldsymbol{b}$, there are two conditions that diffracting planes must satisfy:

$$\boldsymbol{g} \cdot \boldsymbol{b} = 0 \qquad \text{(S1.1)}$$
$$\boldsymbol{g} \cdot (\boldsymbol{b} \times \boldsymbol{u}) = 0 \qquad \text{(S1.2)}$$

where $\boldsymbol{g}$ is the diffraction vector, normal to the set of $hkl$ diffracting planes (in reciprocal space $|\boldsymbol{g}| = 1/d_{hkl}$) and $\boldsymbol{u}$ is the dislocation line direction. By taking dark field images with different $\boldsymbol{g}$ vectors, it is therefore possible to determine the Burgers vector of a dislocation by noting when it becomes invisible.

To obtain the resolution needed to distinguish dislocations in a closely spaced array (as in this work), it is necessary to tilt away from the Bragg condition, resulting in very weak diffracted intensities. The low intensity problem can be circumvented by choosing a diffraction condition that couples the beam of interest to a stronger one, usually in the same systematic row, in a technique known as Weak Beam Dark Field (WBDF) imaging (see Ch. 27 in D. B. Williams, C. B. Carter, Transmission electron microscopy: a textbook for materials science, Springer, New York; London, 2009). In WBDF images, dislocations appear as narrow bright lines against a dark background; the background is dark because the crystal is not near the diffraction condition and the dislocation image is narrow because a large strain field is required to bend the planes back, close to the Bragg condition. This only happens in the high strain field regions next to the dislocation cores.

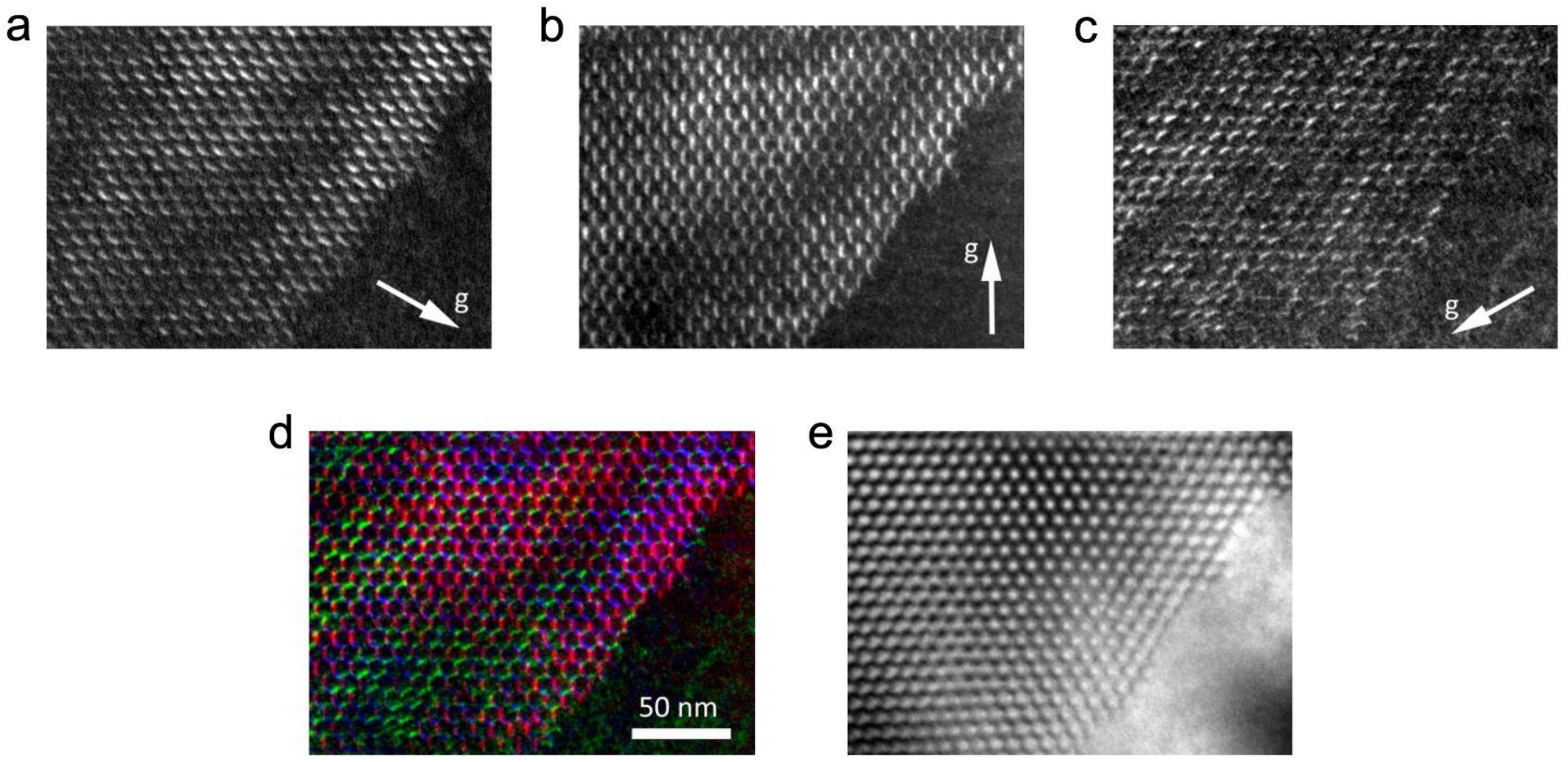

**Figure S1.1: Weak Beam Dark Field (WBDF) Imaging of a Twisted Bonded $LiNbO_3$ Interface**| WBDF images, using different $3\bar{3}00$ $\boldsymbol{g}$ vectors (a)-(c), clearly show bright dislocation strain contrast parallel to each $\boldsymbol{g}$ vector used. The line trace of each dislocation is hence parallel to its Burgers vector, indicating a network of screw dislocations, revealed in the false-coloured superposed image in panel (d). This honeycomb network encloses isolated regions of elastically strained coherently bonded $LiNbO_3$, associated with the bright dot array in the bright-field image using a [0001] zone axis in (e). All images are at the same scale; the 50nm scale bar is indicated in panel d.

In figure S1.1, panels (a) to (c) show WBDF images of a ~58° twisted bonded $LiNbO_3$ interface, obtained using three different $3\bar{3}00$ type diffraction vectors. Here, it is apparent that dislocations form short segments, which only have strong contrast where $\boldsymbol{g}$ is parallel to $\boldsymbol{u}$. From equation (S1.2) above, it is therefore apparent that these are screw dislocations, with Burgers vectors parallel to $\langle 10\bar{1}0 \rangle$ directions. By superposing the three images [fig. S1.1(d)], it becomes clear that the interfacial dislocations exist in a hexagonal honeycomb network. Moreover, by noting the [0001] zone axis bright field image associated with the area in which WBDF imaging was performed, shown in fig. S1.1(e), and with reference to arguments made in the main text, it is also clear that the screw dislocations surround regions of coherent bonding in which torsional elastic strain becomes manifest (and is responsible for the vortical contrast in Differential Phase Contrast and 4D STEM images).

## S2: Moiré Patterns, Coincident Site Lattices and Dislocation Networks

Small crystal misorientation

When two crystals are rotated relative to one another, the superposition of their interfacial lattices produces a moiré pattern. For a small relative rotation $\delta\theta$ from an equivalent orientation, the hexagonal lattices develop a long-period modulation of their relative atomic arrangement, as shown in fig. S2.1(a). The lattice misorientation is accommodated by a periodic network of screw dislocations at the interface [D. Hull and D. J. Bacon, Introduction to Dislocations, 5th ed. (Butterworth-Heinemann, Oxford, 2011); A.P. Sutton and R.W. Balluffi, Interfaces in Crystalline Materials (Clarendon Press, Oxford, 1995); J.P. Hirth and J. Lothe, Theory of Dislocations, 2nd ed. (Wiley, New York, 1982)], as shown in fig. S2.1(b,c). The regions between dislocations remain approximately coherently bonded, while the dislocations accommodate the relative rotation of the crystals. The Frank–Bilby relation connects the misorientation angle $\delta\theta$ to the characteristic dislocation spacing $L_m$. In the small-mismatch limit,

$$|\delta\theta| \simeq \frac{b}{L_m} \tag{S2.1}$$

where $b$ is the magnitude of the Burgers vector. The dislocation network sets the long-period scale, while a smooth elastic reconstruction within each cell may further modulate the local displacement, as shown in fig. S2.1(b), without changing the net misorientation.

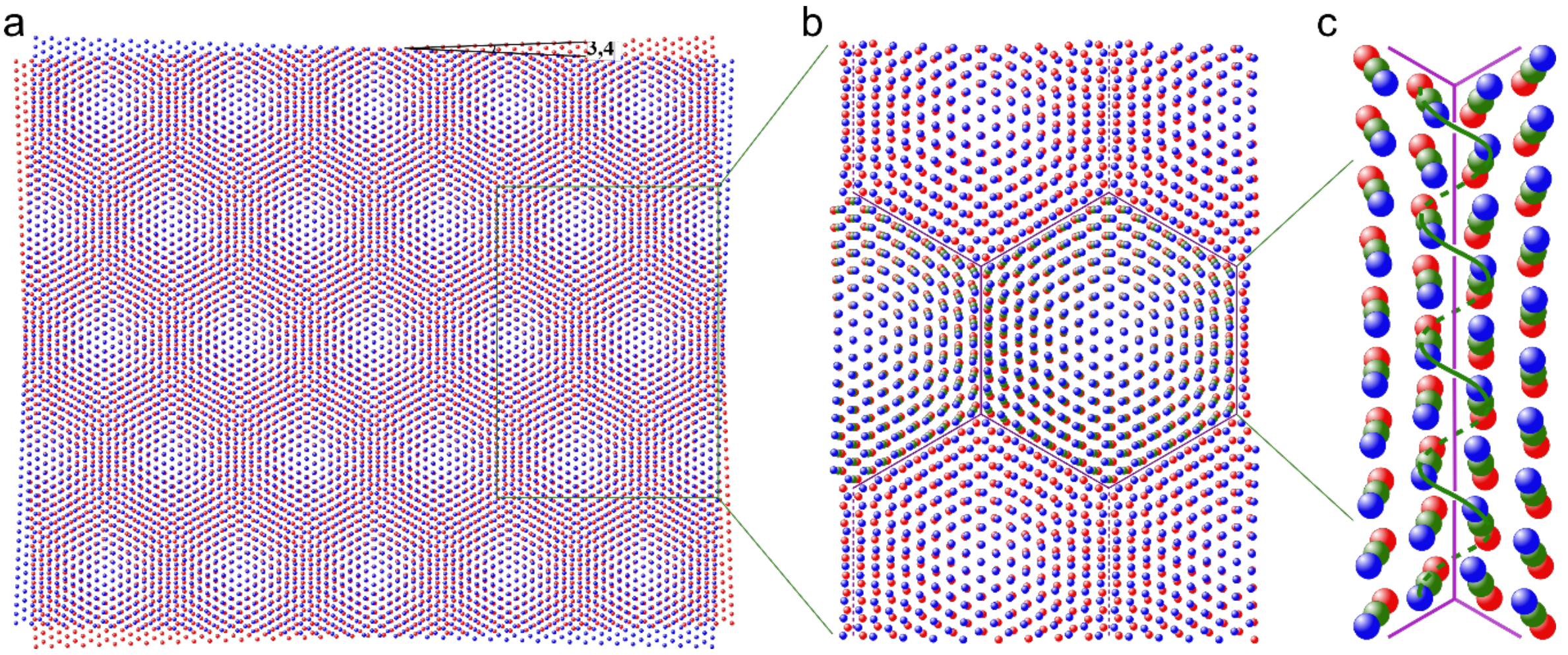

**Figure S2.1: Moiré Structure at the Interface between two Twisted $LiNbO_3$ Crystals Viewed from above |** The upper crystal is rotated clockwise by 4° relative to the equivalent 60° orientation of the lower crystal. (a) Superposition of the undeformed crystal structures near the interface; the interfacial layers of the upper and lower crystals are shown in blue and red, respectively. (b) Enlarged view of the interfacial region after relaxation of the interlayer bonds (green). Magenta lines show the hexagonal network of screw dislocations, and curved arrows indicate the in-plane displacement of the upper crystal layers. (c) Enlarged view of the atoms and interlayer bonds near a screw dislocation at the boundary between adjacent hexagonal cells.

*For the experimentally studied 56° twisted interface* between two $LiNbO_3$ crystals, the nearby commensurate orientation is 60°, which is equivalent to 0° for the (0001) hexagonal mesh. The residual misorientation is therefore $\delta\theta$ = 60° − 56° = 4° = 0.0698 rad. For the interface shown in fig. S2.1(c), the relevant interfacial Burgers vector has magnitude $b = a_H/\sqrt{3} \simeq 0.30$ nm, where $a_H$ is the side length of the hexagonal unit cell. Equation (S2.1) then gives $L_m \simeq b/\delta\theta \simeq$ 4.3 nm. This estimate is consistent with, but slightly smaller than, the experimentally seen period of ~5 nm. The difference may reflect deviations from an ideal pure-twist network, including nonuniform dislocation spacing and weak tilt or edge components [E. Tochigi, Y. Kezuka, N. Shibata, A. Nakamura, Y. Ikuhara, Structure of screw dislocations in a (0001)/[0001] low-angle twist grain boundary of alumina ($\alpha$-$\mathrm{Al_2O_3}$), *Acta Materialia* **60**, 1293–1299 (2012); L.A. Hughes and K. van Benthem, Low-angle twist grain boundary in $SrTiO_3$ fabricated by spark plasma sintering techniques, *Journal of the American Ceramic Society* **102**, 578–586 (2019)].

Large crystal rotation

For a hexagonal lattice, particular relative rotations can produce a coincidence-site lattice (CSL), in which some lattice sites of the two rotated crystals coincide exactly. These commensurate rotations are indexed by coprime integers $m$ and $n$. The corresponding coincidence index, $\Sigma = m^2 + mn + n^2$, gives the ratio between the area of the CSL unit cell and that of a primitive cell of either lattice. The rotation angle is given by $cos\,\theta = (m^2 + 4mn + n^2)/2\Sigma$. The lowest non-trivial solution, $(m, n)$ = (2, 1), gives $\Sigma = 7$ with the $\Sigma 7$ orientation angle $\theta_{\Sigma 7}$ = 21.7868°. In this case, the CSL unit cell has seven times the area of a primitive lattice cell, and one lattice site in every seven is shared by the two rotated lattices (see fig. S2.2(a)).

When the actual rotation slightly deviates from an exact CSL orientation, the coincidence pattern is disrupted, and the resulting mismatch can be accommodated by secondary (sec) grain-boundary dislocations. For a hexagonal CSL with coincidence index $\Sigma$, the shortest relevant Burgers vector has magnitude $b_{\mathrm{sec}} = a/\sqrt{\Sigma}$, where $a$ is the basal-lattice parameter. The dislocation spacing then follows the Frank–Bilby relation, $L_{\mathrm{sec}} \simeq b_{\mathrm{sec}}/|\delta\theta|$, where $\delta\theta$ is the deviation from the exact CSL angle. As illustrated in Fig. S2.2(b), this secondary network sets a long-period scale that can be much larger than the conventional moiré period. Similar secondary dislocation networks [C.P. Sun and R.W. Balluffi, Secondary grain boundary dislocations in [001] twist boundaries in MgO I. Intrinsic structures, *Philosophical Magazine* A **46**, 49–62 (1982)].

*For the experimentally studied 21° twisted interface* between two $LiNbO_3$ crystals, TEM reveals a conventional moiré period of approximately 0.8 nm, whereas the structural and dipolar helices have a much larger period of ~15 nm. The experimental rotation, $\theta$ = 21°, is close to the $\Sigma 7$ orientation (fig. S2.2(a)), with residual misorientation $|\delta\theta|$ = |21° − 21.7868°| ≃ 0.79° = 0.0137 rad. Taking the shortest Displacement Shift Complete (DSC) Burgers vector, $b_{\mathrm{DSC}} = a/\sqrt{7} \simeq$ 0.195 nm, the Frank–Bilby relation gives $L_m \simeq b_{\mathrm{DSC}}/|\delta\theta| \simeq$ 14 nm, close to the observed helix period of ~15 nm, that is consistent with the scale set by the secondary dislocation network (fig. S2.2(b)).

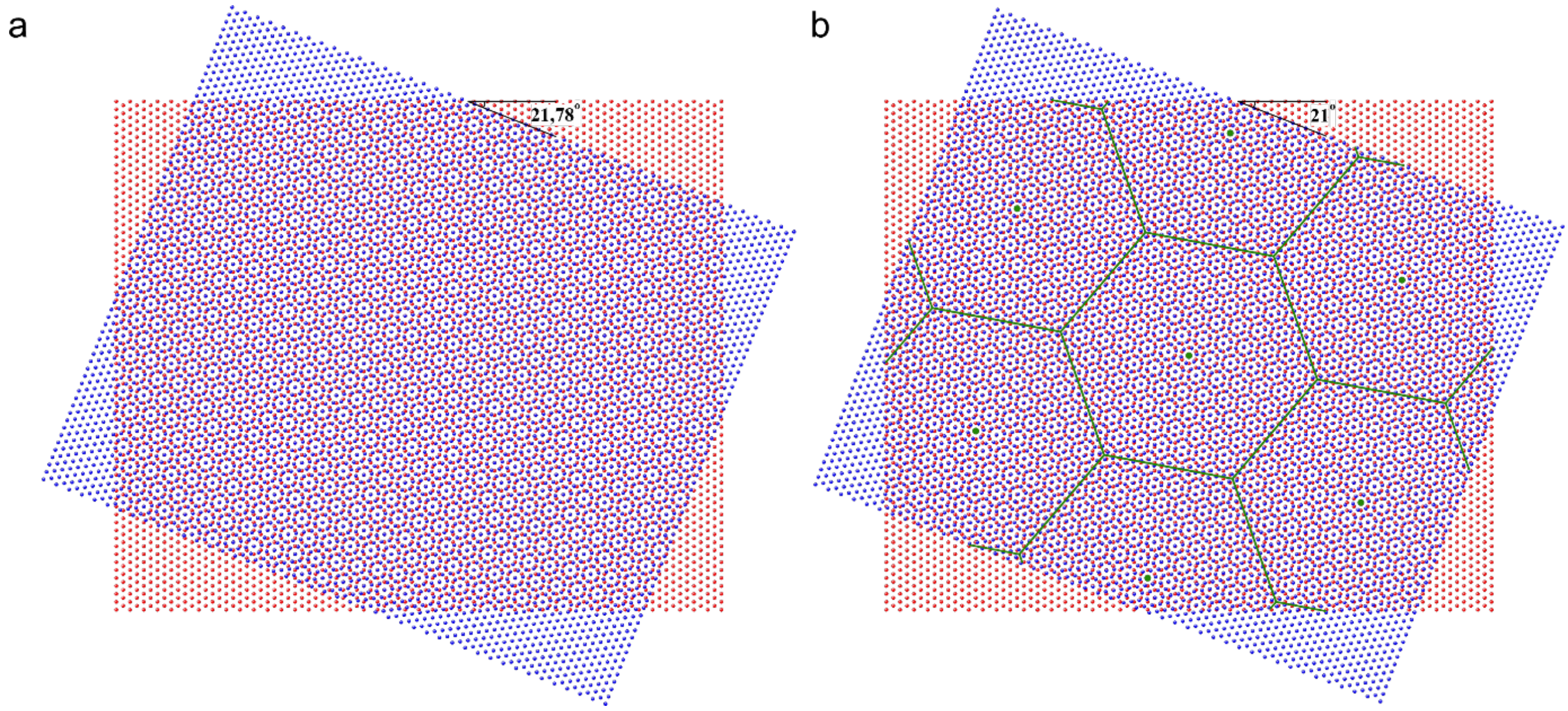

**Figure S2.2: Moiré Structures at Large Twist Angles at the Interface between Two $LiNbO_3$ Crystals |** (a) Exact Σ7 CSL orientation, with a relative rotation of $\theta_{\Sigma 7}$ = 21.78°. The two lattices form a regular periodic coincidence pattern. (b) Relative rotation of $\theta$ = 21°, deviating by approximately $\delta\theta$ = 0.79° from the Σ7 orientation. The resulting mismatch produces a network of secondary dislocations, indicated by the green lines.

## S3: Topological Reconstruction from DPC and 4D STEM Data

The experimentally measured in-plane vector field $\mathbf{V}(\mathbf{r})$, as indicated by arrows in fig. 1e and fig. 2a of the main text and derived from the intensity centroid shifts in 4D-STEM / DPC diffraction patterns, is sampled only at discrete points and may be affected by noise. We continuously interpolate the measured field and use its integral curves, $d\boldsymbol{r}/ds = \mathbf{V}[\mathbf{r}(s)]$, to reveal the detailed field structure shown in fig. S3.1. The singular points act as topological markers that organize the surrounding trajectories and capture essential features. We identify three types: outgoing foci $F_+$, incoming foci $F_-$, and saddle points $S$, with Poincaré indices +1, +1, and −1, respectively [J. Milnor, Topology from the Differentiable Viewpoint, Princeton University Press, 1997].

Based on this analysis, in addition to the dominant vortices at the centres of the hexagonal cells, reconstruction resolves smaller vortices within the vortex blocks that rotate in the opposite direction. Their counter-rotation compensates the rotation of the dominant vortices, so that the field has a rich local rotational structure but no net rotation over a periodic cell. The vortices also have a radial component near the vortex centres: the trajectories spiral outward from F+ and inward toward F−. It should be noted that some of these features are revealed through analysis of the electron ptychography results (particularly compare features in S3.1 with those in figs. S7.1-S7.3).

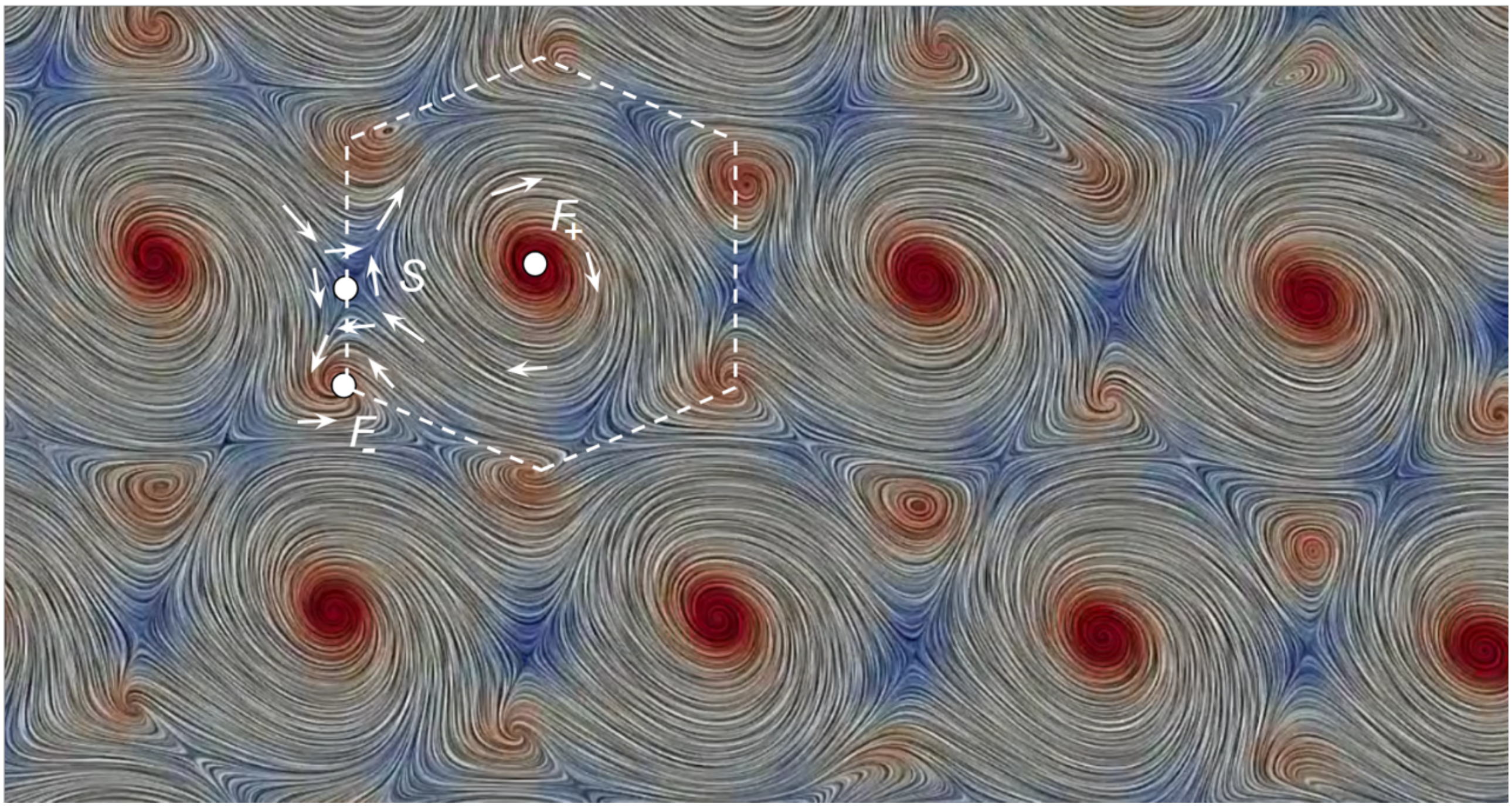

**Figure S3.1: Topological Reconstruction of the Experimental 4D-STEM / DPC Vector Field |** Integral curves of the measured in-plane centre-of-integrated-intensity displacement field. Outgoing foci $F_+$, incoming foci $F_-$, and saddle points $S$, together with the local field direction, are shown in and around an exemplary periodic cell. Colors indicate the degree of local winding.

## S4: 4D STEM Images

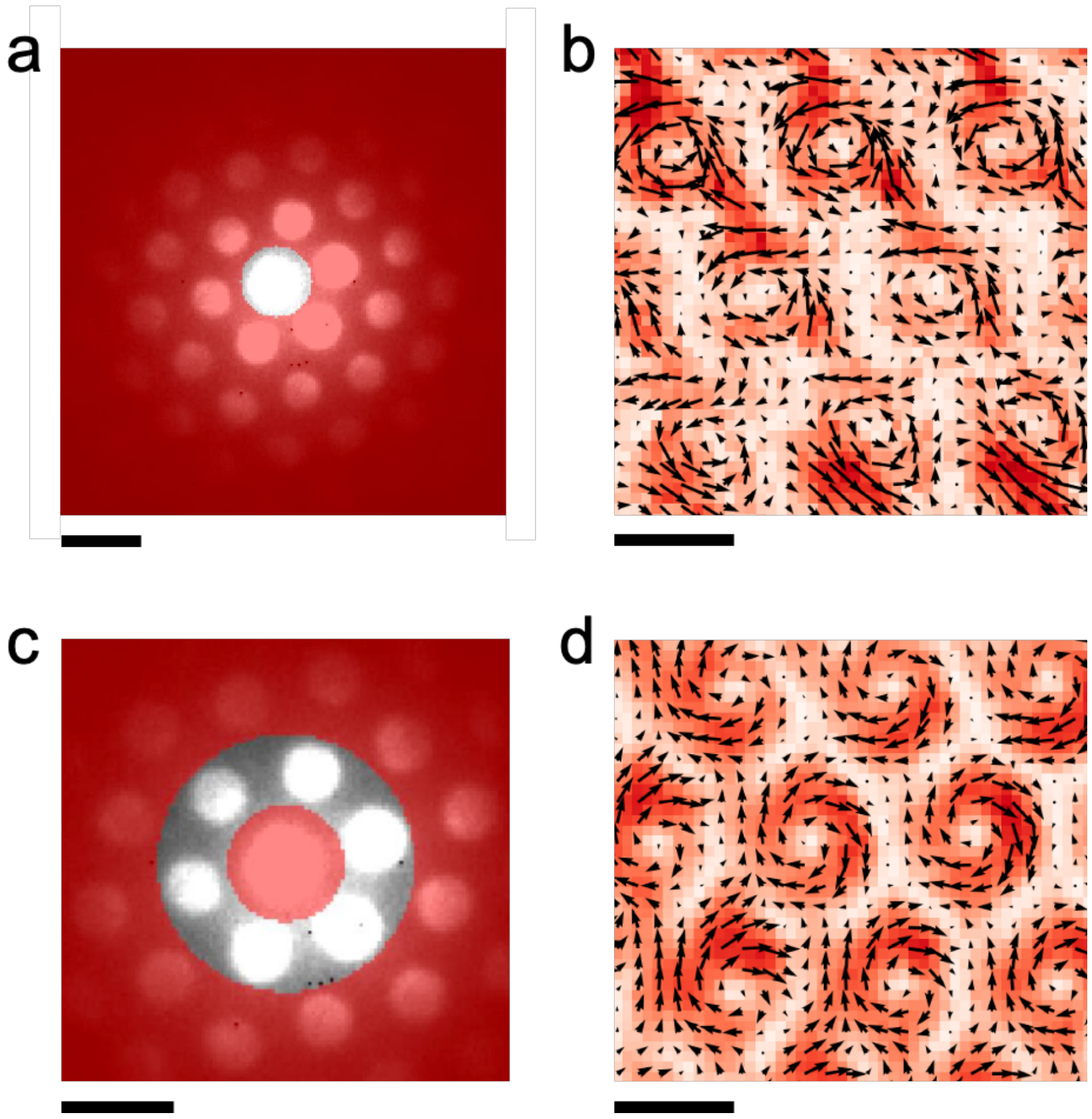

**Figure S4.1: Vorticity implied by 4D STEM and Centre of Integrated Intensity Analysis |** Panel (a) shows a back focal plane diffraction image taken using 4D STEM. When only the straight-through central beam is considered (using a bright field digital aperture), the shift of the centre of the integrated intensity away from the geometric centre of the aperture at each STEM pixel shows a periodic array of anticlockwise vortices (b) with the same periodicity as that associated with the dislocation network. Using a dark-field digital annular aperture instead (c) generates shifts in the centre of integrated intensity that also indicate vorticity, but in the opposite sense (d). Scale bars in a and c are 0.5Å$^{-1}$ and in b and d are 5nm.

## S5: Superposed 4D STEM Vorticity and Strain Information from High Resolution TEM:

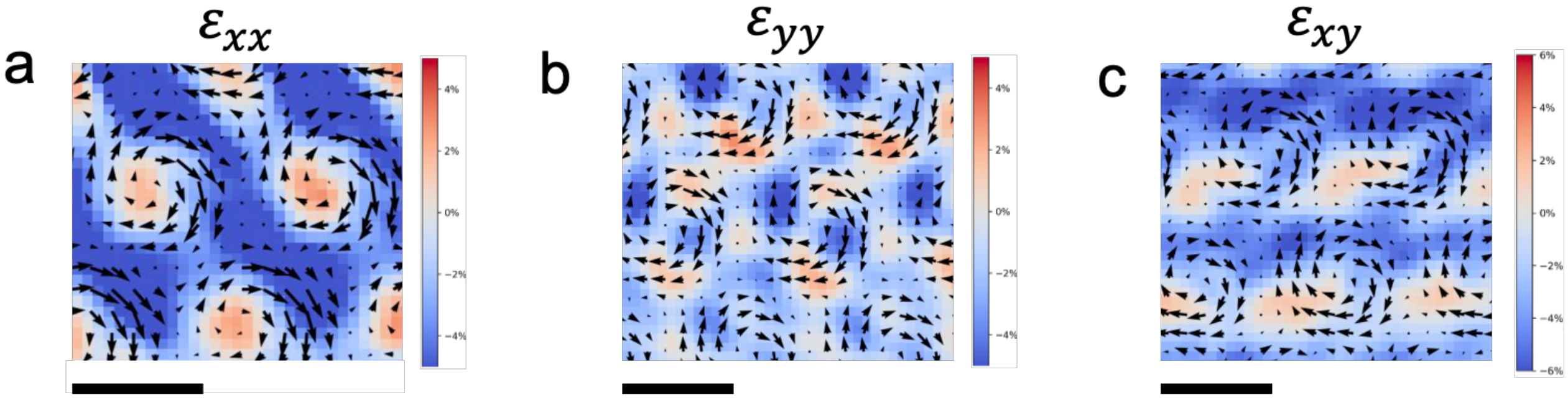

**Figure S5.1: Vorticity spatially correlated with strain analysis |** While the figure in the main manuscript illustrates the spatial superposition of the rigid lattice mesh rotation and the vorticity indicated by 4D STEM, here we show similar maps but with individual strain tensor matrix components considered: diagonal terms (a) and (b) and shear terms (c). All have the same periodicity as the apparent 4D STEM vortices, but the patterns are not as intuitively linked to elastic rotation as the representation in the main manuscript. All scale bars represent 5nm.

## S6: Modelled Elastic Deformation

The screw-dislocation network described above accommodates the relative twist and divides the interface into coherently bonded cells. Within each cell, the lattice undergoes a smooth elastic deformation: interfacial bonds tilt, producing a periodic displacement that extends into the adjoining crystals. The dislocations set the characteristic cell size, while this elastic displacement describes the local lattice rotation within each coherent hexagonal cell without changing the macroscopic misorientation. We represent the in-plane displacement at the interface as a first-harmonic hexagonal modulation of the upper half-space at z = 0 (at the interface and into the upper crystal):

$$u_{\parallel}^{(0)}(x,y) = \sum_{\alpha=1}^{3} [u_l \hat{\mathbf{q}}_\alpha + u_t(\hat{\mathbf{z}} \times \hat{\mathbf{q}}_\alpha)] \sin(\mathbf{q}_\alpha \cdot \mathbf{r}) \tag{S6.1}$$

where $\mathbf{r} = (x,y)$ and the wave vectors $\mathbf{q}_1 = q_m(1,0)$, $\mathbf{q}_2 = q_m(-1/2, \sqrt{3}/2)$, and $\mathbf{q}_3 = q_m(-1/2, -\sqrt{3}/2)$ define the hexagonal star-shaped coherent platelet, with $q_m = 4\pi/(\sqrt{3}L_m)$ and $\hat{\mathbf{q}}_\alpha = \mathbf{q}_\alpha/q_m$. For the symmetric interfacial relaxation considered here, the lower half-space (at the interface and into the lower crystal) undergoes the opposite in-plane displacement, $-u_{\parallel}^{(0)}(x,y)$. Here, $u_l$ and $u_t$ are the longitudinal and transverse amplitudes, respectively. The superposition of the longitudinal harmonics produces a predominantly radial displacement pattern within each cell, whereas the transverse harmonics produce a tangential vortex texture.

In Cartesian coordinates, eq. (S6.1) reads

$$u_x^{(0)}(x,y) = -u_t f(x,y) + u_l g(x,y), \qquad u_y^{(0)}(x,y) = u_t g(x,y) + u_l f(x,y), \tag{S6.2}$$

where

$$f(x,y) = \frac{\sqrt{3}}{2}[\sin(\mathbf{q}_2 \cdot \mathbf{r}) - \sin(\mathbf{q}_3 \cdot \mathbf{r})]$$

$$g(x,y) = \sin(\mathbf{q}_1 \cdot \mathbf{r}) - \frac{1}{2}\sin(\mathbf{q}_2 \cdot \mathbf{r}) - \frac{1}{2}\sin(\mathbf{q}_3 \cdot \mathbf{r}) \tag{S6.3}$$

The transverse component counter-rotates the two lattices within each coherently bonded hexagonal cell, with a maximum relative rotation of $3q_m u_t$. In the limit of complete local compensation, this rotation equals the residual misorientation $\delta\theta$. Using $q_m = 4\pi/(\sqrt{3}L_m)$ gives $u_t \simeq (\sqrt{3}/12\pi)L_m\delta\theta$. For the measured period $L_m \simeq 5$ nm and $\delta\theta$ = 4° = 0.0698 rad, this yields $u_t \simeq 16 - 20$ pm. By symmetry, the maximum rotation of either lattice is half the relative rotation, $\omega_{\max} = 3q_m u_t/2 \simeq \delta\theta/2 \simeq 2°$, consistent with the local lattice rotation of approximately 2-3° extracted from 4D STEM strain analysis of the atomic-column images in fig. 2(a) of the main text. The spatially averaged rotation of this periodic mode vanishes, so it does not contribute to the macroscopic crystal misorientation. Guided by the topological reconstruction shown in fig. S3.1, we take the longitudinal component to be $u_l = 2$ pm, approximately an order of magnitude smaller than $u_t$.

The propagation of the interfacial reconstruction into the upper half-space was determined numerically by solving the elastic equilibrium equations $C_{ijkn}\,\partial_j\,\partial_k u_n = 0$, where $C_{ijkn}$ are the elastic stiffness constants. For trigonal $LiNbO_3$, with *z* along the trigonal axis and the in-plane axes chosen according to the crystallographic convention, the nonzero independent components are $C_{1111} = C_{2222} = 203.0$, $C_{1122} = 57.3$, $C_{1133} = C_{2233} = 75.2$, $C_{3333} = 242.4$, $C_{1313} = C_{2323} = 59.5$, $C_{1212} = 72.8$, and $C_{1123} = C_{1312} = -C_{2223} = 8.5$ GPa, with the tensor symmetries $C_{ijkl} = C_{jikl} = C_{ijlk} = C_{klij}$ [Landoldt1981]. Equation (2) prescribes the in-plane boundary displacement, while the normal displacement is free to relax, with $\sigma_{zz} = 0$ at the interface. Both in-plane amplitudes were retained, with $u_t = 20$ pm and $u_l = 2$ pm. The resulting strain field, $u_{ij} = (\partial_i u_j + \partial_j u_i)/2$, and its spatial decay are shown in fig. 6.1.

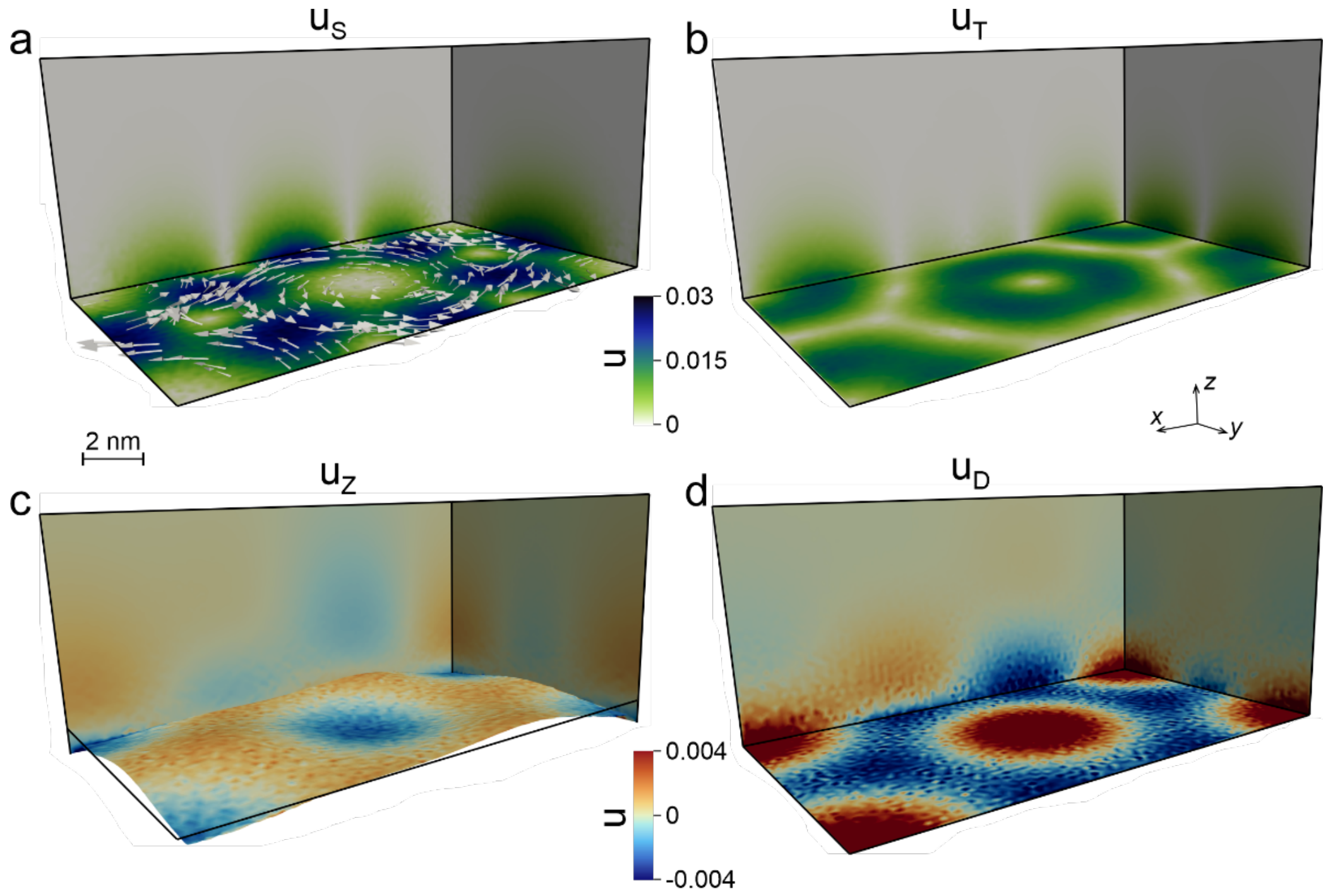

**Figure 6.1: Spatial Distributions of the Strain |** (a) in-plane shear strain, $u_S = \left[(u_{xx} - u_{yy})^2 + 4u_{xy}^2\right]^{1/2}$; white arrows show the in-plane displacement field, dominated by the transverse component $u_t \simeq 20$ pm with a small, $\simeq 10\%$, longitudinal admixture $u_l$; (b) out-of-plane shear strain associated with torsional deformation, $u_T = (u_{xz}^2 + u_{yz}^2)^{1/2}$; (c) dilatational strain along *z*, $u_Z = u_{zz}$; the interface deformation is shown with a characteristic displacement of ±2 pm; (d) in-plane dilatational strain, $u_D = u_{xx} + u_{yy}$. The distributions result from the solution of the anisotropic elasticity problem.

The equation of linear elasticity $C_{ijkn}\,\partial_j\,\partial_k u_n = 0$ is solved using a finite element method. The computational domain is a box with dimensions of 9×8×5 nm, partitioned into unstructured tetrahedra. The discrete solution space for the unknown displacement vector $u$ consists of continuous, piecewise linear Lagrange polynomials and constrained periodically in $xy$-plane. Expressions (S6.2) are used to impose Dirichlet boundary conditions at the lower face of the domain. The assembled system is solved with a generalized minimal residual method and generalized additive Schwarz preconditioner.

The elastic response is presented in fig. S6.1, in terms of four symmetry-adapted strain combinations that are invariant under rotations of the in-plane coordinate axes. Panel (a) shows the in-plane shear amplitude $\mathrm{u_S} = \left[\left(u_{xx} - u_{yy}\right)^2 + 4u_{xy}^2\right]^{1/2}$, which quantifies the local change of shape, excluding areal expansion or compression. The superposed displacement vectors reveal a predominantly vortical in-plane texture. The weak radial component associated with $u_l$ gives this texture a slight spiral character. Panel (b) shows the out-of-plane shear amplitude $\mathrm{u_T} = \left(u_{xz}^2 + u_{yz}^2\right)^{1/2}$, which characterizes the torsional deformation. Panel (c) presents the normal strain $\mathrm{u_Z} = u_{zz}$ together with the corresponding vertical corrugation of the interface. This corrugation could, in principle, be resolved experimentally. Finally, panel (d) shows the in-plane dilatational strain $\mathrm{u_D} = u_{xx} + u_{yy}$, which measures the local areal expansion or compression. The weak normal and dilatational responses arise from the small longitudinal component and anisotropic elastic coupling; both vanish in the simplified isotropic transverse solution considered below.
For simplified analytical calculations, one can neglect the weak longitudinal component by setting $u_l = 0$. The interfacial displacement in eq (S6.2) then reduces to $u_x^{(0)} = -u_t f(x,y)$ and $u_y^{(0)} = u_t g(x,y)$. In the isotropic limit, this transverse mode decouples from the normal displacement, so that $u_z^{(0)} = 0$ and the interface remains planar. Its penetration into the upper half-space is

$$u_x(x,y,z) = -u_t f(x,y)e^{-q_m z}, \quad u_y(x,y,z) = u_t g(x,y)e^{-q_m z}, \quad u_z(x,y,z) = 0. \tag{S6.4}$$

## S7: Determining the Form of the Structural Helices using Electron Ptychography

Helical features of the twisted bonded interface between two z-cut $LiNbO_3$ single crystals are convincingly revealed by electron ptychography, performed on Focused-Ion Beam prepared lamellae, cut at a small angle of inclination with respect to the interface. Ptychographic slices parallel to the (0001) planes, at different distances from the interface, are presented in figures S7.1-S7.3 below. The helical elastic distortions in regions bounded by screw dislocations are explicitly evidenced by the relative displacements of the Nb positions between successive ptychographic slices. The average magnitudes of these displacements are presented as a function of distance away from the interface in figure S7.4. Clearly, the helical pitch (wavelength) changes and tightens to a minimum at the interface itself. The relation evident in figure S7.4 is reminiscent of the derivative of a tanh(ax) function, which is common in domain walls and interfaces and consistent with standard Landau-Ginzburg treatments.

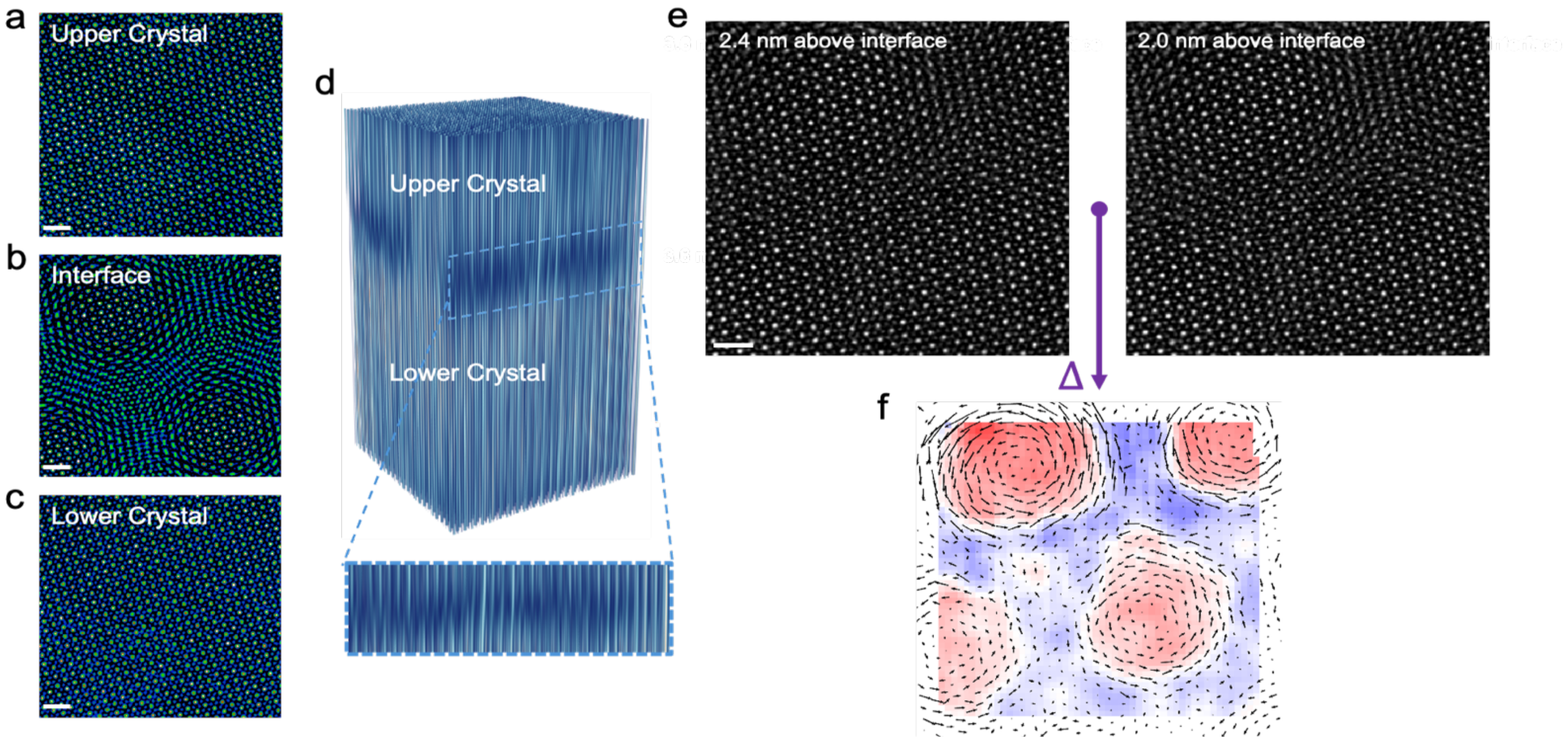

**Figure S7.1 | Multislice Ptychographic Reconstruction of the Twisted $LiNbO_3$ Bilayer.** (a) Phase image summed over the reconstructed slices of the upper LNO crystal. (b) Phase image summed over the interfacial region. (c) Phase image summed over the reconstructed slices of the lower crystal. (d) Three-dimensional rendering of the reconstructed phase contrast. A signature of the periodic disruption in the structure at the interface is clear. (e) By considering two successive ptychographic slices (in this case centred at 2.4nm and 2.0nm above the interface), the change in the Nb positions can be tracked and the displacement vectors plotted (f). Here, the black arrows represent the Nb displacement vectors (accurate in orientation and exaggerated in magnitude). The colours (red and blue) indicate the local toroidal moment associated with the Nb displacement vectors that surround each pixel, within a 7x7 pixel local area. It gives similar information to the local curl of the vector field. Periodic vortical patterns are obvious, even several nanometres away from the interface.

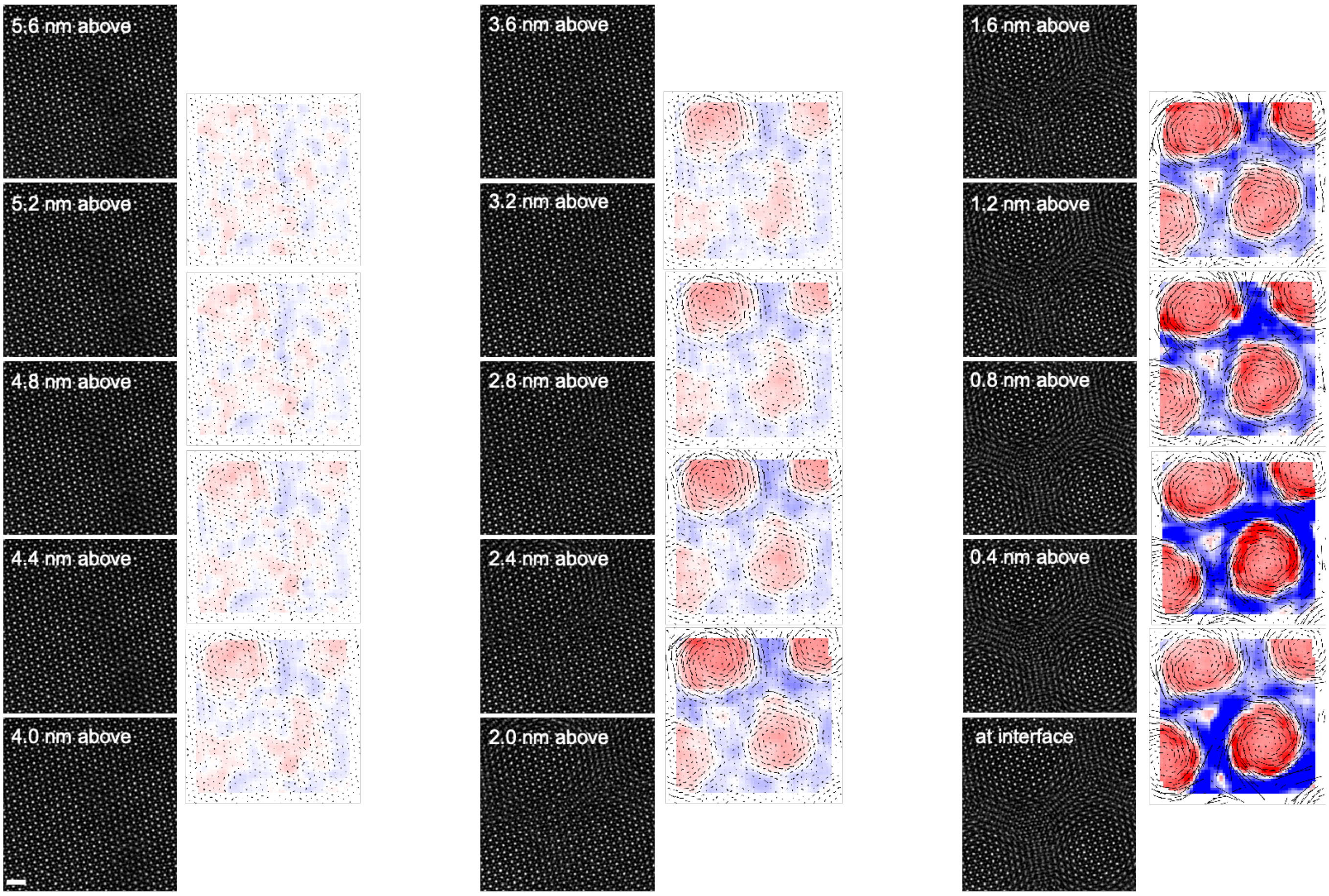

**Figure S7.2 Electron Ptychographic Slices and Displacements of Nb atoms Above the Twisted Interface |** The sets of images in each column above (left, middle, right) present ptychographic reconstructions in the upper crystal, centred at different distances away from the bonded twisted $LiNbO_3$ interface. Alongside these black and white ptychographic slices are coloured maps of the local toroidal moment associated with displacements in the Nb positions on moving from one ptychographic slice to the next (represented by the black arrows). The Nb atom displacements show periodic strong vorticity close to the interface. As can be seen, these features persist for ~4-5nm away from the interface. Importantly, Nb displacements show vortical displacements between a significant number of successive ptychographic slices and thus the loci of their positions in three dimensions are helices. Scale bar (bottom left) represents 1nm.

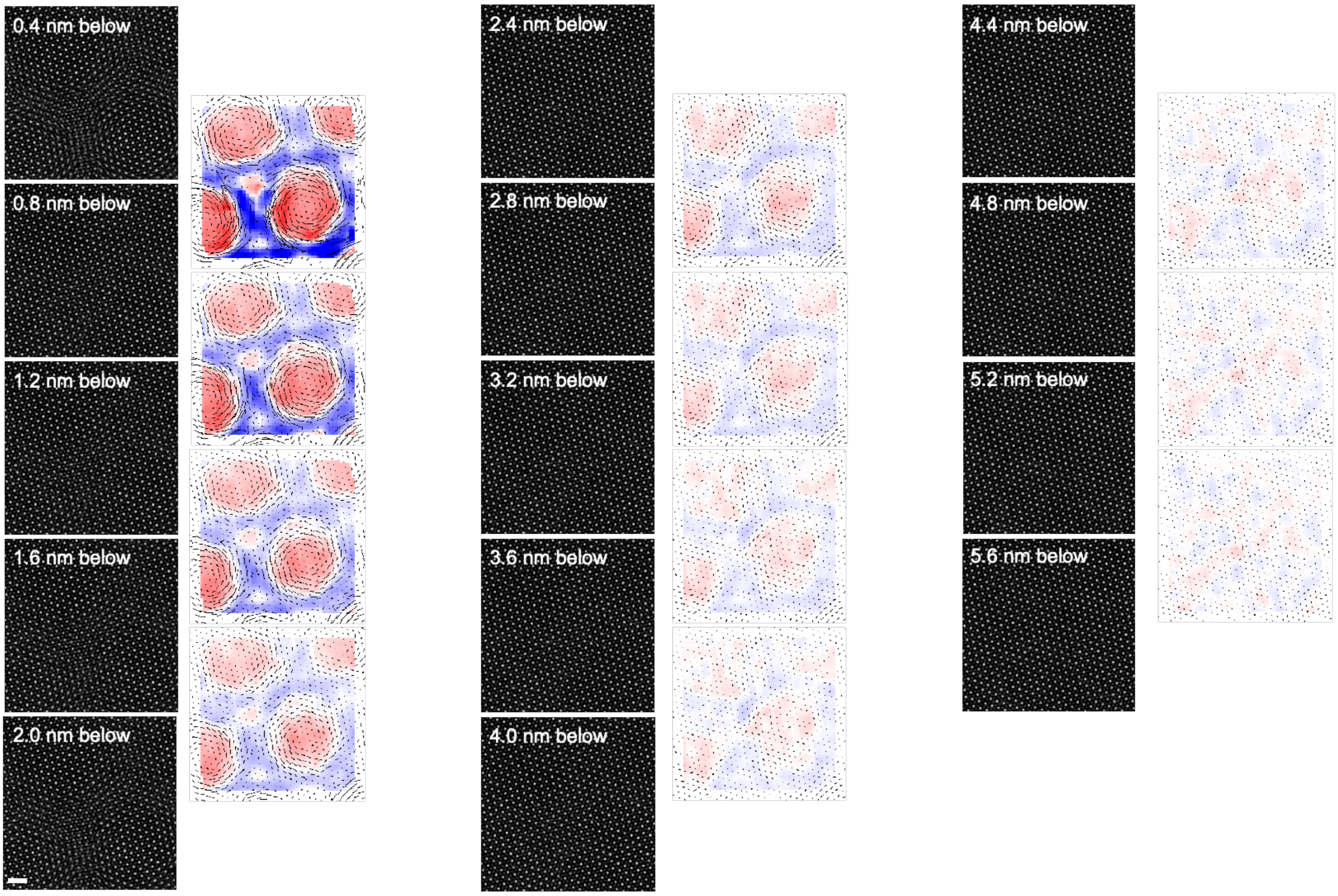

**Figure S7.3 Electron Ptychographic Slices and Nb atom Displacements Below the Twisted Interface |** The sets of images in each column above (left, middle, right) present ptychographic reconstructions in the lower crystal, centred at different distances away from the bonded twisted $LiNbO_3$ interface. Alongside these black and white ptychographic slices are coloured maps of the local toroidal moment associated with displacements in the Nb positions on moving from one ptychographic slice to the next (represented by the black arrows). The Nb atom displacements show periodic strong vorticity close to the interface. As can be seen, these features persist for ~3-4nm away from the interface. Importantly, Nb displacements show vortical displacements between a significant number of successive ptychographic slices and thus the loci of their positions in three dimensions are helices. Moreover, the chirality of the helices is the same below the interface and above the interface - the interfacial region therefore consists of a hexagonal network of screw dislocations bounding an array of elastically deformed helices. Scale bar (bottom left) represents 1nm.

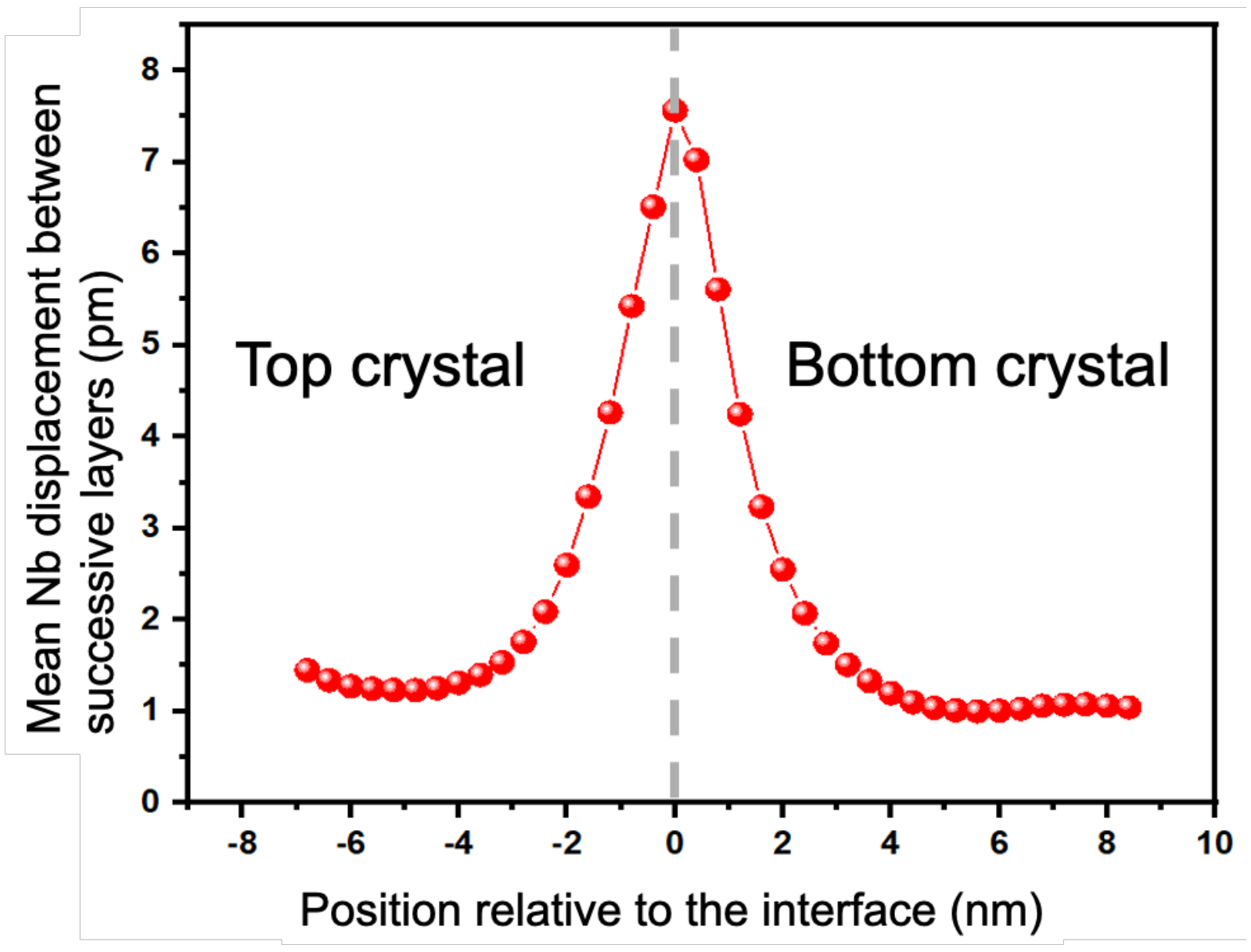

**Figure S7.4 Mean Nb Atomic Displacements between Each Successive Ptychographic Slice as a Function of Distance |** The mean Nb displacement associated with the difference between atomic distributions in successive ptychographic slices demonstrates a clear maximum at the interface. This resembles an expectation of the derivative of a helical twist such that $\theta_x = \theta_{\pm\infty} \tanh\left({}^{x}\!/_{\sqrt{2}\xi}\right)$, where $\theta_x$ is the twist angle at a given displacement $(x)$ from the interface, $\theta_{\pm\infty}$ is the value of twist angle in each of the bulk crystals either side of the interface, and $\xi$ is a scaling parameter.

## S8: Modelling Transport Perpendicular to the Twisted Interfaces (Making Comment on 4D STEM and Differential Phase Contrast)

To pass currents perpendicular to the twisted interface, we use the Hairy Probes method of JPCM **19**, 196201 and PRB **94**, 075118. In this method, we connect **virtual** external probes to the bottom few and the top few layers in the elastically twisted stack. The bottom probes have one electrochemical potential, $\mu_B$, and the top probes have another, $\mu_T$. The electrochemical potential difference $eW = \mu_T - \mu_B$ drives the current.

We adopt the simplest electronic model that tracks the connection between atomistic structure and current flow: an orthogonal single-orbital tight-binding model. We work in the small-bias limit, where the current is linear in $W$. The remaining specification is the Fermi energy $\mu = (\mu_B + \mu_T)/2$, which is a parameter in the calculation.

The calculation uses standard Green's function techniques and works as follows. Each external probe $p$ couples to exactly one atom, $p$, with a coupling strength $\Gamma_p$. The embedding self-energy for the system is then

$$\Sigma = -i\pi \sum_p |p\rangle \Gamma_p \langle p|$$

where $|p\rangle$ is the tight-binding basis state at atom $p$. It is useful, further, to split $\Sigma$ as

$$\Sigma = \Sigma_B + \Sigma_T$$

where the two terms contain the contributions from the bottom and top probes, respectively.

The Green's function for the system then reads

$$G(E) = (E - H - \Sigma)^{-1}$$

where $H$ is the tight-binding Hamiltonian for the stack. The non-equilibrium part of the electronic density matrix is

$$\Delta\rho = -\frac{eW}{2\pi i} G(\mu)(\Sigma_T - \Sigma_B)G^{\dagger}(\mu)$$

Finally, the bond-current between atoms $n$ and $n'$ is

$$I_{nn'} = \frac{2}{\hbar} Im\,[H_{n'n} \Delta\rho_{nn'}]$$

The bond current can be thought of as a vector

$$I_{nn'} = I_{nn'} \hat{R}_{nn'}$$

where $\hat{R}_{nn'}$ is the unit vector joining the two sites. We can associate a net vectorial local current with a given site $n$ as

$$I_n = \frac{1}{2} \sum_{n' \neq n} I_{nn'}$$

The Hamiltonian matrix elements $H_{nn'}$ are functions of interatomic distance. Following Phil. Mag. A **81**, 1833 here they are taken to have inverse power scaling of the form

$$H_{nn'} = t_0 \left(\frac{a}{R_{nn'}}\right)^q$$

where $t_0 = -1$ sets the unit of energy and $a = 1$ sets the unit of length. Physically, $a$ is the nearest-neighbour bond-length in the perfect crystal. In the calculations reported in the main text, $q = 4$.

The Hairy Probes are attached following the graded-probes protocol of PRB **94**, 075118, by turning on the coupling strength $\Gamma_p$ from zero to a chosen maximum value over four layers, at the bottom and top of the stack, respectively. This minimises internal reflections and mimics perfect open boundaries.

The twist for each layer, at vertical position $z$, is given by

$$\theta(z) = \frac{\theta_0}{2} tanh\left(\frac{z - \bar{z}}{\xi}\right)$$

where $\theta_0$ is the total applied twist, $\bar{z}$ is the mid-plane, and $\xi$ controls how many layers the twist is distributed over. When $\xi \to 0$ the twist occurs at one single interface, across the mid-plane of the stack. In the present calculations, $\xi = 2\Delta z$, where $\Delta z = 0.81a$ is the chosen vertical separation of the atomic layers. The atomic layers themselves are hexagonal lattices with AA stacking in the untwisted geometry.

To bring out visually the effect of the twist on the current density (figure 2c in the main manuscript), we have chosen an artificially large total twist $\theta_0 = 35$ degrees. With the above value of $\xi$, this twist is distributed over roughly three neighbouring interfaces. We stress however that the qualitative features of the twist-induced current vorticity are independent of details of the model; they are, rather, an inevitable and general consequence of the chiral bond-tilting induced by the twist.

**S9: Inferred Dipolar Patterns in Elastically Distorted Interfacial Regions**

We consider an *x-y* interface between oppositely polarized $LiNbO_3$ crystals. The uniaxial polarization profile across the tail-to-tail or head-to-head junction is approximated as $\mathbf{P}_0(z) = \pm\hat{\mathbf{z}}P_s \tanh(z/\sqrt{2}\xi_0)$, where $P_s \simeq 0.75\ \mathrm{C\,m^{-2}}$ and $\xi_0 = 0.22\ \mathrm{nm}$ [D. Lee, et al., Structure and energetics of ferroelectric domain walls in $LiNbO_3$ from atomic-level simulations, *Phys. Rev*. B **82**, 014104 (2010)]. The total polarization is written as the sum of this uniaxial profile and the strain-induced piezoelectric and flexoelectric contributions:

$$\mathbf{P} = \mathbf{P}_0 + \mathbf{p}^{\mathrm{piezo}} + \mathbf{p}^{\mathrm{flexo}},$$

$$p_i^{\mathrm{piezo}} = \zeta(z)e_{ijk}u_{jk}, \qquad p_i^{\mathrm{flexo}} = \mu_{ijkl}\,\partial_l u_{jk}. \tag{S9.1}$$

where $\zeta(z) = P_z(z)/P_s$ approximates the reversal of the piezoelectric response across the wall. Since the strain field decays strongly away from the interface, both the piezoelectric and flexoelectric contributions are localized near the interface and vanish in the bulk. Figure S9.1(a) shows the resulting total polarization in the upper half-space (upper crystal) for the head-to-head configuration. The streamlines illustrate how the polarization bends as it approaches the interface due to the piezoelectric and flexoelectric contributions. Arrows at the interface plane indicate the in-plane projection of the polarization vector, and the color map shows the magnitude of this projection. The polarization components in the lower half-space reverse sign.

The point group $C_{3v}$ of $LiNbO_3$ permits four independent piezoelectric constants and ten independent flexoelectric constants. The room-temperature piezoelectric constants are $e_{113} = e_{223} = 3.7$, $e_{222} = -e_{112} = -e_{211} = 2.5$, $e_{311} = e_{322} = 0.23$, and $e_{333} = 1.3\ \mathrm{C\,m^{-2}}$, with $e_{ijk} = e_{ikj}$; all other components vanish [T. Mitsui and S. Nomura, Ferroelectrics and Related Substances, subvol. a: Oxides, in Landolt–Börnstein Numerical Data and Functional Relationships, Springer ed, Vol.16 (1981)]. To the best of our knowledge, the flexoelectric coefficients $\mu_{ijkl}$ have not yet been reported. We therefore use its general isotropic representation in terms of the displacement field $\mathbf{u}$:

$$\mathbf{p}^{\mathrm{flexo}} = \mu_L\nabla(\nabla\cdot\mathbf{u}) - \mu_R\nabla\times(\nabla\times\mathbf{u}). \tag{S9.2}$$

The first term describes the response to longitudinal deformation, whereas the second describes the response to transverse deformation. For further simplification, we set $\mu_L = 2\mu_R = \mu$, which gives $p_i^{\mathrm{flexo}} = \mu\,\partial_j u_{ij}$. We estimate $\mu = \varepsilon_0\bar{\varepsilon}f \simeq 3.5\times10^{-9}\ \mathrm{C\,m^{-1}}$, using the flexocoupling voltage *f* = 10 V adopted in $LiNbO_3$ modeling [S. Cho et al., Switchable tribology of ferroelectrics, *Nat. Commun*. **15**, 387 (2024)] and the orientationally averaged permittivity $\bar{\varepsilon} \simeq 40$. This estimate is consistent with reported flexoelectric coefficients for $SrTiO_3$ and $BaTiO_3$ [P. Zubko et al., *Phys. Rev. Lett*. **99**, 167601 (2007); W. Ma and L.E. Cross, *Appl. Phys. Lett*. **88**, 232902 (2006)].

We decompose the strain-induced polarization correction into piezoelectric and flexoelectric contributions, shown separately in figs. 9.1(b) and (c). The piezoelectric contribution is predominantly in-plane and localized just above the interface, decaying further away with the strain field and changing its orientation as the hexagonal plate helically rotates. Its circulation is opposite to that of the displacement field for the tail-to-tail configuration and has the same sense for the head-to-head configuration. The flexoelectric contribution is also concentrated near the interface and has both in-plane and out-of-plane components. Its pattern depends strongly on the flexoelectric tensor, whose components are not well established for $LiNbO_3$; fig. 9.1 panel (c) therefore shows an illustrative configuration. Our estimate suggests that the flexoelectric contribution is about an order of magnitude smaller than the piezoelectric contribution. It may modify the local polarization pattern, including the weak radial in-plane component associated with the displacement field, but it is likely to be difficult to resolve experimentally against the much larger piezoelectric contribution.

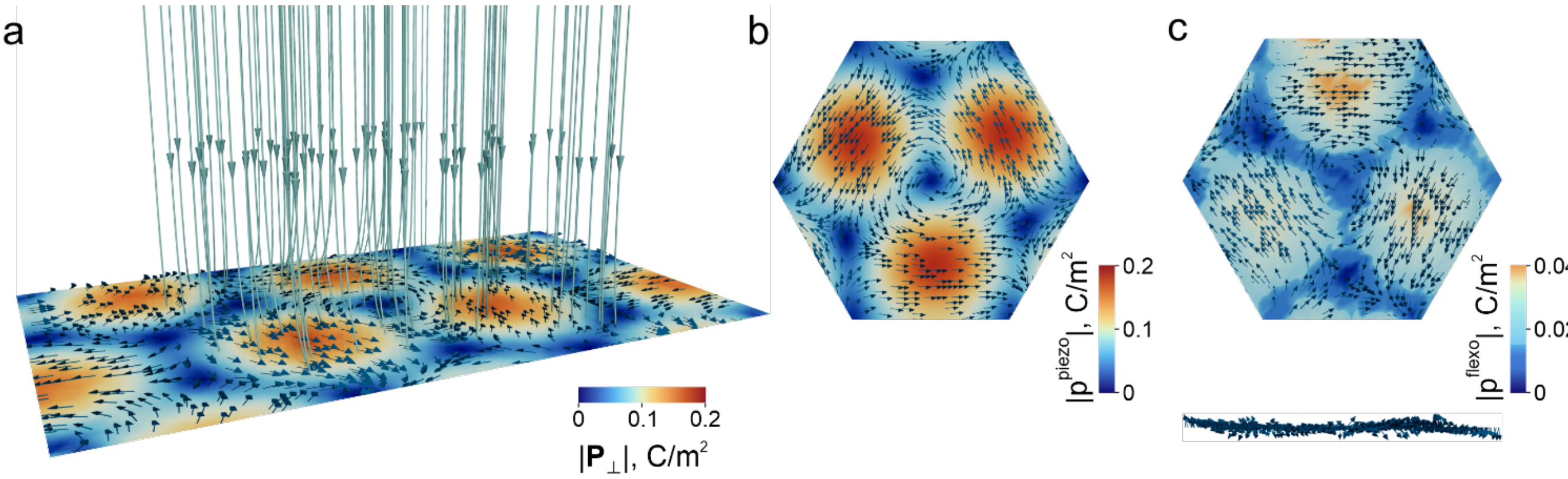

**Figure S9.1: Polarization Profile and its Piezoelectric and Flexoelectric Contributions near the Interface**. (a) Three-dimensional view of the total polarization in the upper half-space for the head-to-head configuration at the interface. The streamlines show how the polarization bends as it approaches the interface due to the piezoelectric and flexoelectric effects. At the interface, arrows indicate the in-plane projection of the bent polarization vector, $\mathbf{P}_{\perp}$, and the color map shows its magnitude. In the lower half-space, all polarization components reverse sign. (b) Piezoelectric contribution in the upper half-space in the vicinity of the interface, viewed from above. Arrows show its in-plane projection, and the color map shows its magnitude. The corresponding out-of-plane component is negligible. (c) Flexoelectric contribution in the upper half-space in the vicinity of the interface. The arrows in the upper image show its in-plane component, viewed from above, and the color map shows its magnitude. The lower image is a side view showing the out-of-plane flexoelectric component.

## S10: Geometry-Dependent Circular Dichroic Second Harmonic Generation (CD-SHG): Identification of a Sample-Linked Pseudoscalar Contribution

To distinguish sample-linked circular dichroism from contributions fixed in the optical apparatus, CD-SHG measurements were performed in two specimen configurations, denoted geometries A and B. Geometry B was obtained from geometry A by rotating the specimen by 180° in the sample plane about the surface normal. The optical arrangement itself was left unchanged: the propagation direction, focusing and collection optics, incident polarisation states, analyser, and detector all remained fixed in the laboratory frame. Consequently, instrumental effects that depend only on this fixed laboratory geometry are expected to be invariant under the A→B operation. Residual ellipticity, imperfect generation of circular polarisation, birefringence or phase shifts in optical components, helicity-dependent transmission, and detector imbalance are all fixed in the laboratory frame and therefore remain unchanged when only the sample is rotated. Such effects may contribute to the geometry-even background but cannot, on their own, generate a reproducible geometry-odd term.

The A→B experiment shows that the odd component of the dichroic signal requires a sample-dependent nonlinear response whose projection onto the fixed optical frame changes under specimen rotation. The A→B comparison provides a direct control against a purely instrumental origin of the sign-reversing signal. The transformation behaviour of the odd contribution is consistent with a pseudoscalar nonlinear response associated with the interfacial chiral state. For a helical polarisation texture, the natural pseudoscalar is the polarisation-helicity density $H_P = P \cdot (\nabla \times P)$. Here P is a polar vector, whereas $\nabla \times P$ is an axial vector; their scalar product is therefore a pseudoscalar, changing sign under spatial inversion.

In this study, electron microscopy establishes the existence of structurally helical interfacial distortions, and the elastic and piezoelectric analysis predicts transverse polarisation components coupled to this handed strain field. Their superposition on the longitudinal ferroelectric polarisation provides a microscopic route to a helical dipolar texture and hence to finite polarisation helicity. The structural helices can therefore be viewed as a handed elastic template: they provide the geometrical and symmetry environment in which a transverse polarisation texture can acquire a preferred helicity. The polar discontinuity adds a strong electrostatic boundary condition that can generate, stabilise or enhance transverse interfacial polarisation. Coupling of this transverse component to the handed structural distortion provides a natural route to a finite pseudoscalar $H_P$.

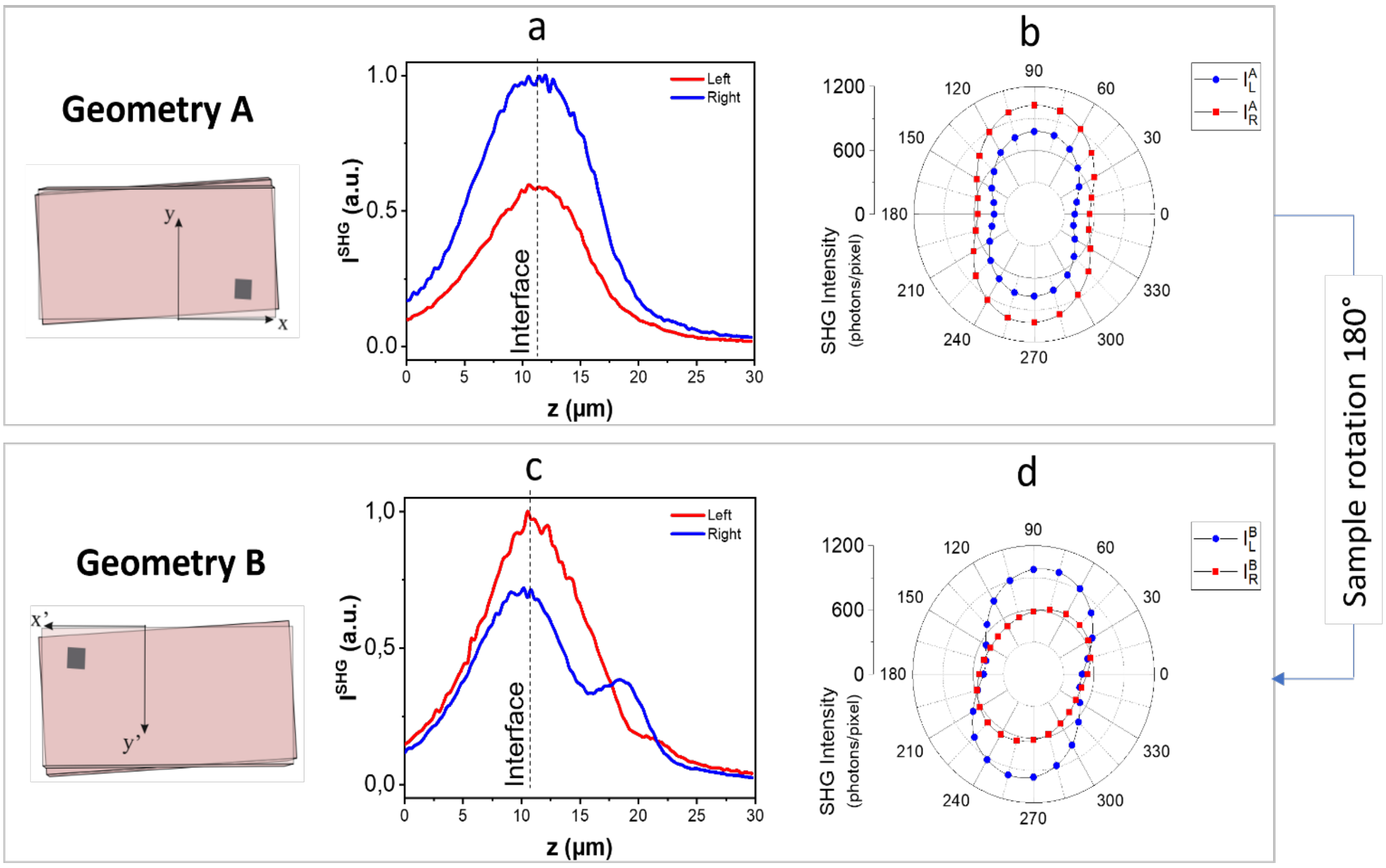

**Figure S10.1: Geometry-Dependent CD-SHG Response Consistent with a Polarization-Helicity Pseudoscalar |** In geometry A, the SHG profile across the interface (a) and the corresponding polarization dependence (b) show a higher SHG intensity under right-circularly polarized excitation. After rotation of the sample to geometry B, the SHG profile (c) and polarization dependence (d) instead show a higher intensity under left-circularly polarized excitation. The resulting reversal of the CD-SHG contrast demonstrates that the sign of the dichroic response is linked to the sample orientation while the laboratory optical frame remains unchanged. When the sample is rotated by 180° in-plane, its intrinsic chirality, and hence the sign of the polarisation-helicity density, remain unchanged. However, the projection of the sample-linked nonlinear response onto the fixed optical frame changes, producing the observed reversal of the geometry-odd CD-SHG contribution.